\documentclass[11pt]{article}
\usepackage[utf8]{inputenc}
\usepackage{subfiles}
\usepackage{amsmath}
\usepackage{xcolor}
\usepackage{graphicx}
\usepackage{algorithm} 
\usepackage{algpseudocode} 
\usepackage{setspace}
\usepackage{hyperref}
\usepackage[symbol]{footmisc}

\usepackage{authblk}

\usepackage{geometry}
\usepackage{natbib}
\usepackage{multibib}
\newcites{sm}{References for Supplementary Material}

\graphicspath{ {./figures/} }

\title{\textbf{Conditional canonical correlation estimation based on covariates with random forests} \vspace{0.6cm}}
\author[ ]{Cansu Alakuş$^{1,}$\thanks{Corresponding author. E-mail: \href{mailto:cansu.alakus@hec.ca}{cansu.alakus@hec.ca}}}
\author[ ]{Denis Larocque$^{1,}$\thanks{After the first author, A Labbe and D Larocque are (equally) the main contributors of this article.}}
\author[3,4]{S\'ebastien Jacquemont}
\author[4]{Fanny Barlaam}
\author[4]{Charles-Olivier Martin}
\author[4]{Kristian Agbogba}
\author[2,4]{Sarah Lipp\'e}
\author[1,$\dagger$]{Aur\'elie Labbe}
\affil[1]{Department of Decision Sciences, HEC Montr\'eal, Montr\'eal, QC H3T 2A7, Canada} 
\affil[2]{Department of Psychology, Universit\'e de Montr\'eal, Montr\'eal, QC H3T 1J4, Canada}
\affil[3]{Department of Pediatrics, Universit\'e de Montr\'eal, Montr\'eal, QC H3T 1C5, Canada}
\affil[4]{CHU Sainte-Justine Research Center, Universit\'e de Montr\'eal, Montr\'eal, QC H3T 1C5, Canada}
\date{}

\begin{document}

\maketitle
\begin{abstract}
    Investigating the relationships between two sets of variables helps to understand their interactions and  can be done with canonical correlation analysis (CCA). However, the correlation between the two sets can sometimes depend on a third set of covariates, often subject-related ones such as age, gender, or other clinical measures. In this case, applying CCA to the whole population is not optimal and methods to estimate conditional CCA, given the covariates, can be useful. We propose a new method called Random Forest with Canonical Correlation Analysis (\texttt{RFCCA}) to estimate the conditional canonical correlations between two sets of variables given subject-related covariates. The individual trees in the forest are built with a splitting rule specifically designed to partition the data to maximize the canonical correlation heterogeneity between child nodes. We also propose a  significance test to detect the global effect of the covariates on the relationship between two sets of variables. The performance of the proposed method and the global significance test is evaluated through simulation studies that show it provides accurate canonical correlation estimations and well-controlled Type-1 error. We also show an application of the proposed method with EEG data.
\end{abstract}

\newpage
\section{Introduction} \label{sec:Introduction}

Data from multiple sources, called multi-view data, refers to many types of data that include complementary information from different aspects to characterize a subject. For example, in biomedical studies, the collection of data may include subject-related covariates (e.g. age, gender, medical history), DNA sequencing, transcriptomics (e.g. mRNA, microRNA, RNA sequencing) and proteomics for a single subject \citep{cancer2012comprehensive, encode2012integrated}. As another example, in functional neuroimaging, we may have subject-related covariates (e.g. age, gender, intracranial volume (ICV)), brain imaging data, and cognitive measurements for subjects \citep{fratello2017multi}. Integration of multiple feature sets and investigating the relationships between them may help to understand their interactions and obtain more meaningful interpretations. Studying the integration of multiple feature sets requires statistical and machine-learning tools which include methods for dimension reduction, clustering, classification, and association studies for multi-view data integration (see \cite{sun2013survey, meng2016dimension, min2017deep, li2018review} for comprehensive reviews).

Canonical correlation analysis (CCA), firstly introduced in \cite{hotelling1936}, is a multivariate statistical method that analyzes the relationship between two multivariate data sets, $X$ and $Y$. CCA searches for linear combinations of each of the two data sets, $Xa$ and $Yb$, having maximum correlation. In CCA, the components $Xa$ and $Yb$ are called canonical variates and their correlations are the canonical correlations. CCA is a two-view data integration tool. It was later generalized to data with more than two views \citep{kettenring1971canonical}. There are some extensions of CCA for under-determined data sets through regularized CCA \citep{vinod1976canonical,cruz2014fast}, for sparse data sets through sparse-CCA \citep{witten2009penalized,hardoon2011sparse} and for nonlinear relationships through generalized CCA \citep{melzer2001nonlinear}, deep CCA \citep{andrew2013deep}, kernel CCA (KCCA) \citep{Akaho01akernel} and non-parametric CCA (NCCA) \citep{michaeli2016nonparametric}. Although very flexible, all these methods suppose that the relationship between the two sets of variables is constant for all subjects, which is not always the case in practice. For example, hundred of gene-environment studies have shown that gene effects on diseases are modulated by environmental factors \citep{caspi2006gene,hunter2005gene,ma2011varying}. In the field of neuroscience, age and gender are known to  interact with brain-behaviour correlations \citep{davis2008pasa,li2010gender} and this should be accounted for in the analyses. In this paper, we focus on CCA and specifically on extending it to account for a subject-related covariate effect on the correlation.

\begin{figure}
    \centerline{\includegraphics[width=0.7\textwidth]{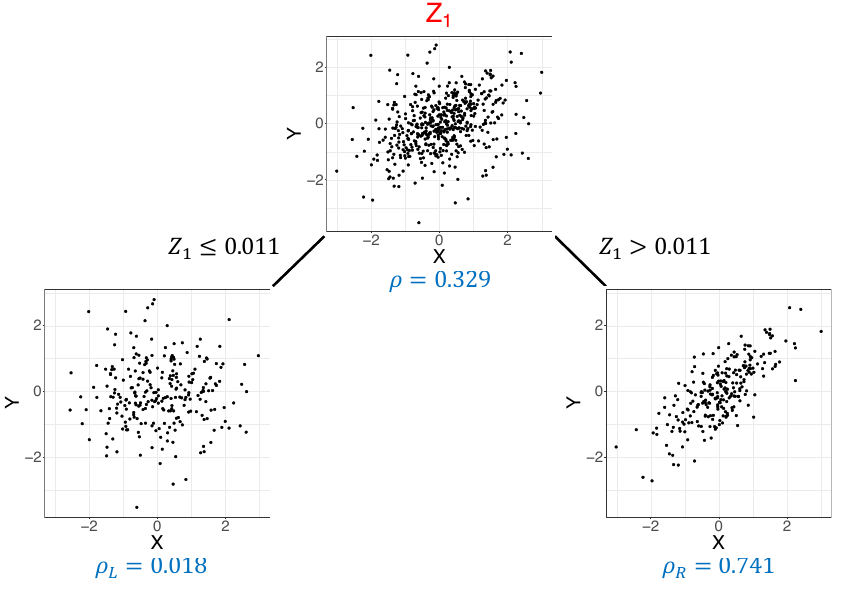}}
    \caption{Example of a dataset $(X,Y)$ where the correlation between $X$ and $Y$ depends on a covariate $Z_1$. This can be captured by a single split of the decision tree.}
    \label{fig:toyextree}
\end{figure}

There are several ways to account for subject-related covariates in multi-view data integration. They can be treated as one of the views \citep{hanna2010anthropometric,li2017incorporating,moser2018multivariate,mihalik2020multiple} or they can be used to identify subgroups in the data while analyzing the relationships between other views. Recently, \cite{choi2020recursive} proposed a recursive partitioning approach, namely correlation tree, to identify homogeneous correlated subgroups within data. In their work, the data consists of a set of subject-related covariates $Z$ (e.g. age, gender, education) and two univariate continuous variables $X$ and $Y$ which are assumed to be correlated. The correlation tree method grows a decision tree with covariates to identify subgroups of subjects with different correlations between the two univariate variables. A simple illustration of this approach is shown in Figure \ref{fig:toyextree} with a single split of the decision tree. In this example, the overall correlation between $X$ and $Y$ is $\rho=0.329$. However, this hides the fact that the subgroup with $Z_1 > 0.011$ has a much higher correlation of $\rho_R=0.741$ while the other subgroup $Z_1 \leq 0.011$
has almost no correlation ($\rho_L=0.018$). Such a situation can be modeled in practice with a regression model including an interaction effect between $X$ and $Z$, but the situation is not as straightforward when both $Y$ and $X$ are multivariate and when the interaction pattern with $Z$ is complex enough to not be captured efficiently by a single tree. Therefore, the goal of this paper is to propose a novel way to estimate the canonical correlations between two sets of multivariate variables $X$ and $Y$, depending on $Z$, using a random forests framework.

Random forest is an ensemble method which contains many decision trees. It is a powerful prediction method due to its ability to limit over-fitting. Moreover, random forests can be seen as a way to find nearest neighbor observations that are close to the one we want to predict \citep{hothorn2004bagging,lin2006random,moradian2017l1,moradian2019survival,roy2020prediction,tabib2020non}.  Each tree in the proposed random forest framework  is built with a new splitting rule designed to produce child nodes with maximum difference in the canonical correlation between $X$ and $Y$. For a new observation with subject-related covariate values $z^*$, the proposed random forest provides a set of similar observations from the training data set that will be used to compute a canonical correlation estimate given $z^*$. Moreover, we propose a significance test to detect the global effect of $Z$ on the relationship between $X$ and $Y$.

This paper is organized as follows. In Section \ref{sec:method}, we describe the proposed method, global significance test and variable importance measure. In Section \ref{sec:simulation}, simulation study results for accuracy evaluation and global effect of covariates are presented to show the performance of the method. A real data example is provided in Section \ref{sec:realdata}, followed by concluding remarks in Section \ref{sec:conclusion}.

\section{Proposed method} \label{sec:method}

In this section, we describe the proposed method in detail.

\subsection{Canonical correlation analysis (CCA)} \label{subsec:cca}

Canonical correlation analysis (CCA), firstly introduced in \cite{hotelling1936}, seeks vectors of $a \in R^{p}$ and $b \in R^{q}$, given two mean centered multivariate data sets $X \in R^{n \times p}$ and $Y \in R^{n \times q}$, such that $Xa$ and $Yb$ are maximally linearly correlated. We can formulate the problem as finding the coefficients $(a^*, b^*)$ such that:
\begin{equation} \label{eq:maxcca}
    (a^*, b^*) = \operatorname*{argmax}_{a,b} corr(Xa, Yb),
\end{equation}
where
\begin{equation*}
    corr(Xa, Yb) = \frac{a^T \Sigma_{XY} b}{\sqrt{a^T \Sigma_{XX} a} \sqrt{b^T \Sigma_{YY} b}},
\end{equation*}
where $\Sigma_{XX}$ and $\Sigma_{YY}$ are the covariance matrices of $X$ and $Y$, respectively, and $\Sigma_{XY}$ is the cross-covariance matrix. Since rescaling $a$ and $b$ does not affect  $corr(Xa, Yb)$, we can add the constraints $a^T \Sigma_{XX} a = 1$, $b^T \Sigma_{YY} b = 1$ to the maximization problem \eqref{eq:maxcca}. There are several ways to solve the CCA problem, such as solving standard or generalized eigenvalue problems \citep{hotelling1936, hardoon2004canonical, bach2002kernel}, using alternating least squares regression  \citep{branco2005robust,wilms2015sparse} and using singular value decomposition \citep{healy1957rotation, ewerbring1990canonical}. 

\subsection{Tree growing process} \label{subsec:splitcriterion}

We use an unsupervised random forest based on the set of covariates $Z$ to find subgroups of observations with similar canonical correlations between $X$ and $Y$. This random forest consists of many unsupervised decision trees with a specialized splitting criterion. The tree growing process follows the CART approach \citep{breiman1984classification}. The basic idea of tree growing with CART is to select the best split at each parent node among all possible splits to obtain the purest child nodes. Assume we want to split a parent node with $n_P$ observations into two child nodes, namely left and right nodes. To split a node, all possible splits are evaluated with the selected splitting criterion. A split value for each split point is calculated by using the observations in the parent node. Then, the best split is selected among all possible splits. The CART algorithm evaluates all possible splits to decide the split variable and split point. In random forests, instead of evaluating all possible splits, the best split search is limited to a randomly selected subset of covariates. Splitting continues until all nodes become terminal nodes.

Since the goal is to find subgroups of subjects with distinct canonical correlations, we propose a splitting rule that will seek to increase the canonical correlation heterogeneity as fast as possible \citep{athey2019generalized, moradian2017l1, tabib2020non}. Define $\rho_L$ and $\rho_R$ as the canonical correlation estimations of the left and right nodes, respectively. The proposed splitting criterion is 
\begin{equation} \label{eq:split} 
\sqrt{n_Ln_R}*|\rho_L - \rho_R|, 
\end{equation} 
where $n_L$ and $n_R$ are the left and right node sizes, respectively. The best split among all possible splits is the one that maximizes \eqref{eq:split}. \cite{choi2020recursive} propose a similar criterion without the $\sqrt{n_Ln_R}$ term among one of three possible splitting criteria for their correlation tree method.

\subsection{Random forest and estimation of canonical correlation for new observations} \label{subsec:rf}

The previous section describes the splitting criterion and tree growing process for a single tree. The final canonical correlations are estimated with a random forest. Random forest \citep{breiman2001random} is a data-driven weight generator.
For example, for a continuous outcome, we can represent random forest predictions for a new observation as a weighted average of the true responses such as $\hat{o}^{new} = \sum_{i=1}^{N}\hat{w}_i(c^{new}) o_i$,
where $\hat{w}_i$ are the predicted weights from the random forest, $c^{new}$ are the covariates of the new observation, and $o_i$ are the observed outcomes \citep{hothorn2004bagging,lin2006random}.

A slightly different representation of these weights was presented in \cite{moradian2017l1} and later  used in \cite{moradian2019survival}, \cite{roy2020prediction} and \cite{tabib2020non}. For a new observation, we form a set of observations which includes the training observations that are in the same terminal nodes as the new observation. \cite{roy2020prediction} called this set of observations  the Bag of Observations for Prediction (BOP). We can define BOP for a new observation $c^{new}$ as
\begin{equation*}
    BOP(c^{new}) = \bigcup\limits_{b=1}^{B} S_b^{new},
\end{equation*}
where $S_b^{new}$ is the set of training observations that are in the same node as $c^{new}$ in the $b^{th}$ tree. Any desired measure can be obtained by using the constructed BOP. In this paper, we use the BOP idea to estimate the canonical correlations for the new observations. Once we train the random forest, we can estimate the correlation for any new observation. In our problem, for a new observation with covariates $z^{new}$, we firstly form  $BOP(z^{new})$. Then, we apply canonical correlation analysis for $X$ and $Y$ with the observations in $BOP(z^{new})$, to compute the canonical correlation estimation $\hat \rho(z^{new})$. 

We can estimate CCA components only if the sample size, $n$, is larger than $(p+q)$. In fact, if $n < (p+q)$, the first $(p+q-n)$ canonical correlations will be exactly one and uninformative \citep{pezeshki2004empirical}. When $n > (p+q)$, although we can estimate the CCA correlations, overfitting can still be a problem. As the sample size in proportion to $(p+q)$ increases, the likelihood of overfitting decreases. During the tree building process, the number of observations in the nodes are getting smaller as we move down in the tree. When the sample size of a node in proportion to a fixed total number of $X$ and $Y$ variables is close to one, we are more likely to overfit. Therefore, we need to control the minimum sample size in the nodes and we do it using the \texttt{nodesize} feature of random forests.

\subsection{Global significance test} \label{subsec:globalsignificance}

For each tree grown in the forest, the proposed method uses the covariate space to identify groups of observations with similar canonical correlations. By doing so, we might be tempted to assume that the set of covariates is indeed relevant to distinguish between canonical correlations. However, this might not be the case and we propose a hypothesis test to evaluate the global effect of the covariates on the canonical correlation. The unconditional canonical correlation between $X$ and $Y$ can be found by computing CCA using the whole sample. If there is a global effect of $Z$ on such correlations, the estimated conditional canonical correlations with the proposed method should be significantly different from the unconditional canonical correlation. %between $X$ and $Y$. 
We perform the following statistical significance test for the null hypothesis 
\begin{equation} \label{eq:null}
    H_0 : \rho(X,Y | Z) = \rho_{\tiny CCA}(X,Y),
\end{equation}
where $\rho(X,Y | Z)$ is the conditional canonical correlation between $X$ and $Y$ given $Z$, and $\rho_{\tiny CCA}(X,Y)$ is the unconditional CCA correlation in the population. 

Before describing the global significance test, we will describe how to estimate canonical correlations for the training data using out-of-bag (OOB) observations. We train a random forest with $B$ trees using the training observations. Each tree $b=\{1,..., B\}$ is built with the selected random bootstrap sample, i.e. inbag observations ($IB^b$), that contains approximately 63\% distinct observations from the original sample. The remaining training observations are the OOB observations for that tree, namely $OOB^b$, and they are not used for building the  $b^{th}$ tree. After training a random forest with $B$ trees, we have ($IB^b$,$OOB^b$) sets for each tree. The estimation of canonical correlation with OOB observations is described in Algorithm \ref{alg:oobest} for a training set with a sample of size $n$. Basically, for a given training observation, the canonical correlation is estimated with the BOP, but using only the trees for which that observation is OOB.

\begin{algorithm}[]
	\caption{Estimation of canonical correlation for a training observation $z_i$ with OOB observations} 
	\begin{algorithmic}[1]
    \For {i=1,...,n}
        \For {b=1,...,B}
            \If {$z_i \in OOB^b$}
                \State Find the terminal node of $z_i$ at tree $b$, say $d$
                \State $BOP_{oob}(z_i) = BOP_{oob}(z_i) \cup IB^b_d(z_i)$ (where $IB^b_d(z_i)$ is the inbag observations
that are in the same terminal node $d$ as $z_i$)
            \EndIf
        \EndFor
        \State Apply CCA for $X$ and $Y$ with the observations in $BOP_{oob}(z_i)$ to find the estimated canonical correlation $\hat \rho(z_i)$
    \EndFor
	\end{algorithmic} 
	\label{alg:oobest}
\end{algorithm}

The proposed global significance test is described in Algorithm \ref{alg:global}. Firstly, we apply CCA to all $X$ and $Y$ to compute the unconditional canonical correlation, which is the root node correlation, say $\rho_{root}$. Then, we apply the proposed method for $X$,$Y$,$Z$ and estimate the canonical correlations for each training observation, $\hat \rho(z_i)$, by using OOB observations as described in Algorithm \ref{alg:oobest}. Finally, we compute the global test statistic with

\begin{equation} \label{eq:globaltest}
    T = \frac{1}{n} \sum_{i=1}^{n}{\big(\hat \rho(z_i) - \rho_{root}\big)^2}.
\end{equation}

The global test statistic is the mean squared difference between the unconditional canonical correlation between $X$ and $Y$, and the estimated canonical correlations with the proposed method. It measures how far the estimated canonical correlations are spread out from the unconditional canonical correlation between $X$ and $Y$. The larger $T$ is, the more evidence against $H_0$ we have. We perform a permutation test under the null hypothesis \eqref{eq:null} by randomly permuting rows of $Z$. For each permuted $Z$, we compute the global test statistic  \eqref{eq:globaltest} and estimate a $p$-value with
\begin{equation} \label{eq:pvalue}
    p = \frac{1}{R} \sum_{r=1}^{R}{I(T'_r > T)},
\end{equation}
where $T'_r$ is the test statistic for the $r^{th}$ permuted $Z$ and $R$ is the total number of permutations. If the $p$-value is less than the pre-specified significance level $\alpha$, we reject the null hypothesis \eqref{eq:null}. 

\begin{algorithm}[H]
	\caption{Global permutation test for covariates' effects} 
	\begin{algorithmic}[1]
	\State Compute CCA for $X$ and $Y$ in the root node, say $\rho_{root}$
	\State Train RF with $X$, $Y$, $Z$
	\State Compute estimated canonical correlations with Algorithm \ref{alg:oobest}, say $\hat \rho(z_i)$
	\State Compute test statistic with $T = \frac{1}{n} \sum_{i=1}^{n}{\big(\hat \rho(z_i) - \rho_{root}\big)^2}$
	\For {$r = 1:R$}
	    \State Permute rows of $Z$, say $Z_{r}$
	    \State Train RF with $X$, $Y$, $Z_{r}$
        \State Compute estimated canonical correlations with Algorithm \ref{alg:oobest}, say $\hat \rho'_r(z_i)$
        \State Compute test statistic with $T'_r = \frac{1}{n} \sum_{i=1}^{n}{\big(\hat \rho'_r(z_i) - \rho_{root}\big)^2}$
    \EndFor
    \State Approximate the permutation $p$-value with \eqref{eq:pvalue}
    \State Reject the null hypothesis when $p < \alpha$. Otherwise, do not reject the null hypothesis.
	\end{algorithmic} 
	\label{alg:global}
\end{algorithm}

\subsection{Variable importance} \label{subsec:variableimportance}

Random forests use OOB samples to construct variable importance (VIMP) measures by evaluating the average change in prediction accuracy. However, we do not have a true response variable since the problem is unsupervised by nature. Therefore, we use the predicted values to compute any VIMP measures we want from existing packages. The idea is that we are measuring the importance of the variables by using a regression forest to  reproduce the canonical correlation estimations we obtained from our method. Hence, the VIMP measures reflect the predictive power of the variables on the estimated canonical correlations. Therefore, higher value of VIMP measure implies higher importance for the estimation of canonical correlations. We propose a two-step process to estimate VIMP measures; see Section \ref{supp:vimpexplain} of the Supplementary Material for details. 

\subsection{Implementation}

We utilised the custom splitting feature of the \texttt{randomForestSRC} package \citep{rfsrc} to implement our splitting criterion in the tree building process. We have developed an R package called \texttt{RFCCA}. The package is available on CRAN, \href{https://CRAN.R-project.org/package=RFCCA}{https://CRAN.R-project.org/package=RFCCA}.

\section{Simulations} \label{sec:simulation}

Evaluating the performance of the proposed method with a real data set is not possible since the true relationships between $X$ and $Y$ with varying $Z$ are typically unknown. Hence, we perform a simulation study with a known data generation process (DGP) to show the performance of our method. We first describe the DGP used in the simulation study. Next, we construct scenarios to validate the proposed global significance test. Since we know the true correlations, we can evaluate the accuracy of the correlation estimations. We can also evaluate the estimated importance ranking of the covariates  since we know the set of covariates that are effectively related to the relationship between $X$ and $Y$, and the ones that are redundant. 

\subsection{DGP} \label{subsec:dgp}

Assume we want to generate a data set with $X \in R^{n \times p}$, $Y \in R^{n \times q}$ and $Z \in R^{n \times r}$ where $n$ is the sample size. We firstly generate the covariates $Z$  according to a standard multivariate normal distribution with an equicorrelated covariance matrix, that is from  $N(0,\Sigma_Z)$ where $\Sigma_{Z}^{}=(1-\rho_z)\textbf{I}_{r} + \rho_z\textbf{J}_{r}$, $\textbf{I}_{r}$ and $\textbf{J}_{r}$ are $r\times r$ identity matrix and matrix of ones, respectively. Then, the true correlation between $Xa$ and $Yb$ for each observation $i$, $\rho(z_i)$, is generated with the following logit model 
\begin{equation*} \label{eq:logit}
    \rho (z_i) = \frac{1}{1+ \exp(-(\beta_0^{}+\sum_{l=1}^{r}\beta_{l}^{} z_{il}^{} + z_{i1}^2))},
\end{equation*}
where $\beta_0^{}$ is the intercept parameter and $\beta_l^{}$ are the weights for the $Z$ variables. In the simulations, we set all $\beta_{l}^{}$ to be $1/r$.

Next, we generate the $X$ and $Y$ coefficients for CCA, $a$ and $b$ in \eqref{eq:maxcca} respectively, with the following two linear equations:
\begin{equation*} \label{eq:lin}
\begin{split}
  a_{ij} &= \max \{0, (1 - s_x \times \rho(z_i) \times j)\}  \ \ \forall j=\{1,2,...,p\},\\
  b_{ik} &= \max \{0, (1 - s_y \times \rho(z_i) \times k)\}  \ \ \forall k=\{1,2,...,q\},
\end{split}
\end{equation*}
where $s_x$ and $s_y$ are some slope parameters to be defined. Note that according to these equations, the $X$ and $Y$ variables have a descending order of relative importance. Also, we may set some coefficients to be 0 with an  appropriate choice of $s_x$ and $s_y$, in order to have redundant covariates that are not directly related to the canonical correlations. 

Finally, for each observation $i$, we generate the $X$ and $Y$ variables with a multivariate normal distribution $N(0,\Sigma_i)$, where $\Sigma_i = \left( \begin{matrix} 
\Sigma_{X}^{} & \Sigma_{XY}^i \\
\Sigma_{YX}^i & \Sigma_{Y}^{}\\
\end{matrix} \right)$, $\Sigma_{X}^{}=(1-\rho_x)\textbf{I}_{p} + \rho_x\textbf{J}_{p}$, $\Sigma_{Y}^{}=(1-\rho_y)\textbf{I}_{q} + \rho_y\textbf{J}_{q}$ and $\Sigma_{XY}^i = \rho(z_i) \Sigma_{X}^{} a_{i}^{} b_{i}^T \Sigma_{Y}^{}$. See Section \ref{supp:dgp} of the Supplementary Material for examples of sample distributions with different parameter settings.

\subsection{Simulation design}

\subsubsection{Evaluation of the power of the global significance test}

In order to evaluate the effect of $Z$, we consider four scenarios where two of them are under the null hypothesis \eqref{eq:null} and the other two are under the alternative hypothesis. For all scenarios, we generate $X$ and $Y$ with two levels of CCA correlation defined to be low (0.3) and high (0.6). For the first scenario (case 1 under $H_0$),
these levels represent the population correlation. For the other three scenarios, they represent the mean sample correlation. We generate the data sets for these scenarios as follows:
\begin{enumerate}
    \item $H_0$ (case 1): We generate 5 $X$, 5 $Y$ with a constant population canonical correlation and 10 $Z$ variables which are all independent and following a standard normal distribution. In this case, the correlation between $X$ and $Y$ is independent of $Z$ and we are therefore under the null hypothesis. 
    \item $H_0$ (case 2): We first generate 5 $X$, 5 $Y$ and 5 $Z$ with the proposed DGP. Then, we replace the $Z$ set with 10 independent $Z$ variables generated with a standard normal distribution. In this case, the correlation between $X$ and $Y$ varies with some of the $Z$ variables but those $Z$ variables are not available in the training set. Hence, although the correlation between $X$ and $Y$ is a function of covariates, these covariates are not part of the training set. Therefore, we are again under the null hypothesis. 
    \item $H_1$ (without noise): We generate 5 $X$, 5 $Y$ and 5 $Z$ with the proposed DGP, and the covariates are available in the training set. In this case, the correlation between $X$ and $Y$ varies with all $Z$ variables.
    \item $H_1$ (with noise): We generate 5 $X$, 5 $Y$ and 5 $Z$ with the proposed DGP and we add 5 independent $Z$ variables to the covariates' training set. In this case, the correlation between $X$ and $Y$ varies with some of the $Z$ variables.
\end{enumerate}

Each of the scenarios above are repeated under the low and high correlation settings, which leads to a total of 8 scenarios to investigate. We also investigate how the performance of the proposed method varies with the training sample sizes, and we use  $n_{train}=\{200,300,500,1000,1500\}$. The number of permutations in the permutation test is set to 500 and each scenario is repeated 500 times. Type-1 error is estimated as the proportion of rejection in the scenarios simulated under $H_0$. Similarly, we estimate power as the proportion of rejection in the scenarios simulated under $H_1$.

For each replication of the simulation, we obtain an estimated $p$-value from the permutation test. If the $p$-value is less than the significance level $\alpha=0.05$, we reject the null hypothesis. The proportion of rejection is then calculated over the 500 replications.

\subsubsection{Accuracy evaluation} \label{subsubsec:accuracyevaldesign}

We perform a simulation study to evaluate the accuracy of the method for estimating the canonical correlation. Table \ref{tbl:param} presents the DGP parameter settings for these simulations. We generate $Z$ with $(r,r^{noise}) = \{(1,5),(5,5),(10,5)\}$ where $r^{}$ and $r^{noise}$ are the number of important and noise $Z$ variables, respectively. We generate $X$ and $Y$ with $(p,q) = \{(1,1),(5,5),(10,10)\}$ where $p$ and $q$ are the number of $X$ and $Y$ variables, respectively. We use training sample sizes of $n_{train}=\{100,200,300,500,1000,5000\}$. Also, we consider six values for parameter \texttt{nodesize} that controls the size of the trees: \texttt{nodesize} = $\{2\times (p+q), 3\times (p+q), 4\times (p+q), 6\times (p+q), 8\times (p+q), 10\times (p+q)\}$. Overall, these combinations produce 648 settings (2 mean CCA correlation levels $\times$ 3 $Z$ dimensionality $\times$ 3 $X$ and $Y$ dimensionality $\times$ 6 training sample sizes $\times$ 6 \texttt{nodesize} levels). Each setting is repeated 100 times for a total of 64,800 runs. In each run, we generate an independent test set of new observations with $n_{test} = 1000$. 

\begin{table}[]
\centering
\begin{tabular}{ccc} 
\hline
\textbf{Parameters} & \textbf{Low CCA correlation} & \textbf{High CCA correlation} \\ \hline
$(p, q)$ & (1,1), (5,5), (10,10) & (1,1), (5,5), (10,10) \\
$(r, r^{noise})$ & (1,5), (5,5), (10,5) & (1,5), (5,5), (10,5) \\
$(\rho_x, \rho_y, \rho_z)$ & (0.3, 0.3, 0.1) & (0.3, 0.3, 0.1) \\
$(\beta_0, \beta_l)$ & (-2,$\frac{1}{r}$) & (-0.3,$\frac{1}{r}$) \\
$(s_x, s_y)$ & (0.7, 0.4) & (0.4, 0.3) \\\hline          
\end{tabular}
\caption{DGP parameter settings for accuracy evaluation simulations. We consider two levels of average CCA correlation: low (0.3) and high (0.6).}
\label{tbl:param}
\end{table} 

We evaluate the performance with the mean absolute errors (MAE), given by
\begin{equation} \label{eq:mae}
    MAE =  \frac{1}{n_{test}} \sum_{i=1}^{n_{test}} |\hat \rho(z_i) - \rho(z_i)|,
\end{equation}
where $\hat \rho(z_i)$ is the estimated canonical correlation and $\rho(z_i)$ is the true correlation for the  $i^{th}$ test observation with covariates $z_i$. Smaller values of the $MAE$ show better performance. We use ordinary CCA, without covariates, as a simple benchmark method. In this case, we let $\rho_{train}$ be the training sample estimated canonical correlation with CCA. This value is used as the correlation estimation for all new observations from the test set. 

\subsection{Results}

\subsubsection{Global significance test}

\begin{figure}
    \centerline{\includegraphics[width=0.9\textwidth]{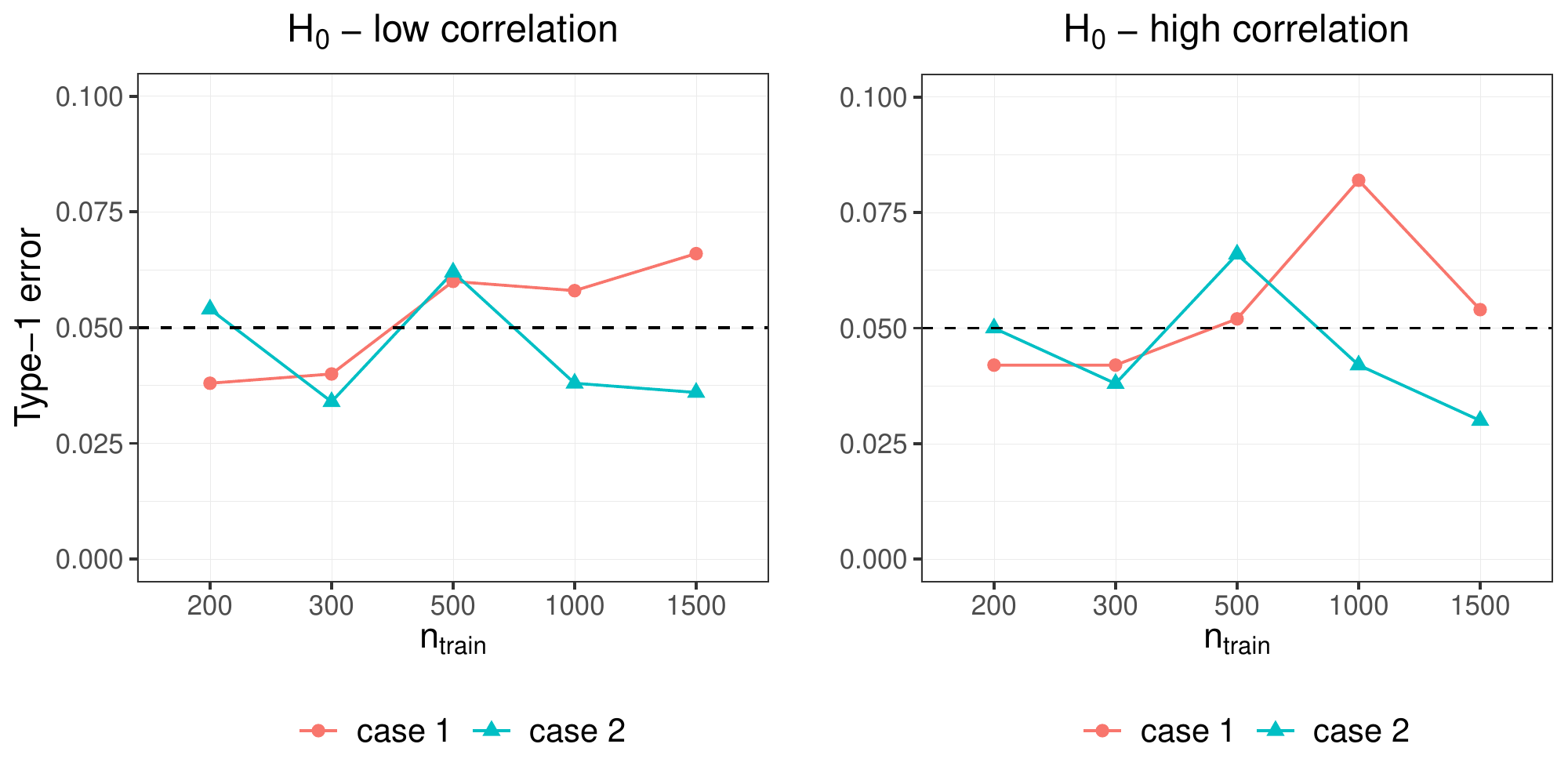}}
    \caption{Global significance test results for $H_0$ cases. In case 1 the correlation between $X$ and $Y$ is not varying with $Z$ and in case 2 the correlation between $X$ and $Y$ is varying with some of the $Z$ variables but those $Z$ variables are not used in the training set. Left and right plots are for low and high correlated data sets, respectively. Dashed line represents the significance level of $\alpha=0.05$.}
    \label{fig:permutation1}
\end{figure}

\begin{figure}
    \centerline{\includegraphics[width=0.9\textwidth]{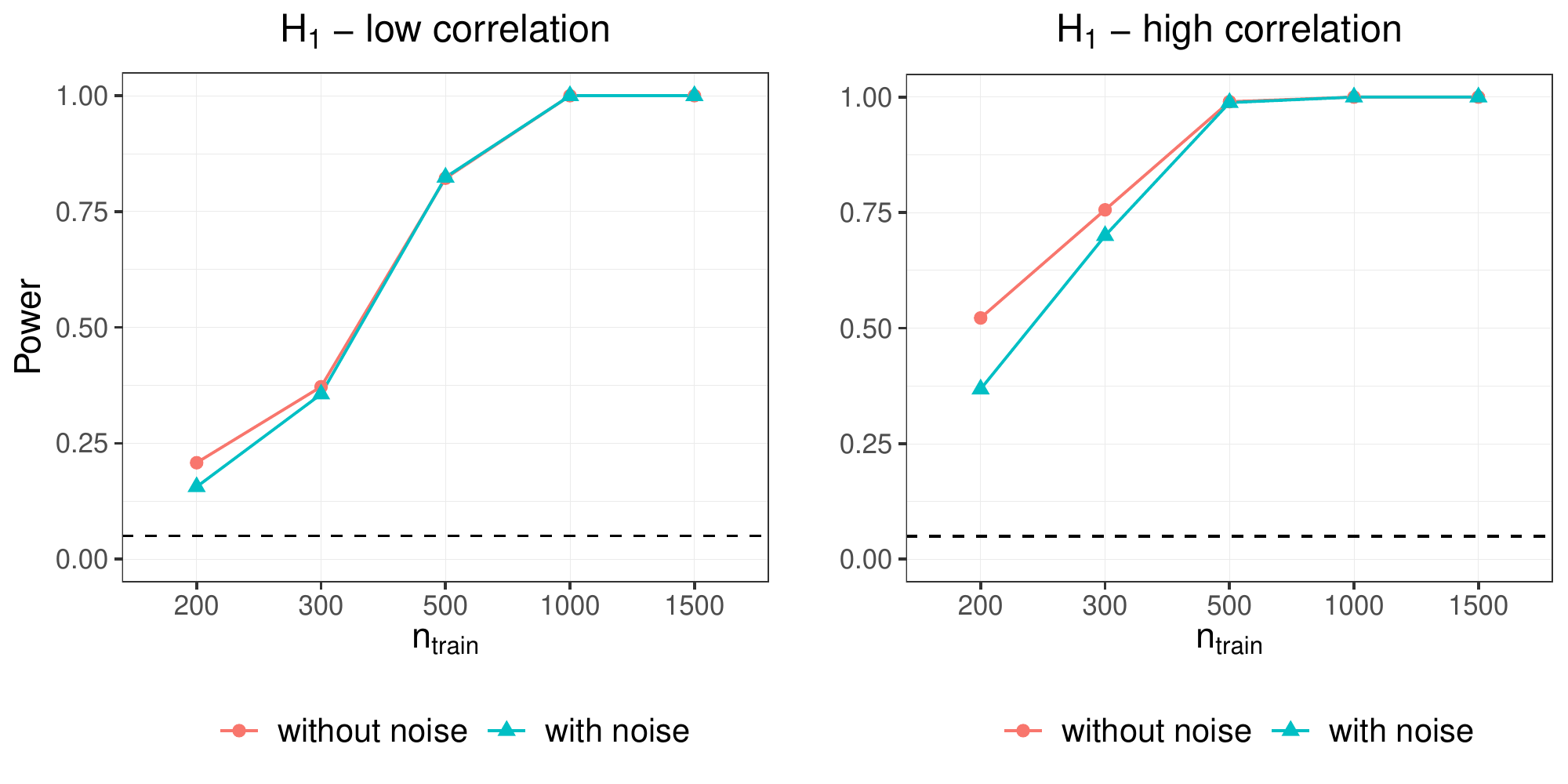}}
    \caption{Global significance test results for $H_1$ scenarios. In without noise case, we have only important covariates for the varying correlation between $X$ and $Y$ and in with noise case we have additional noise covariates. Left and right plots are for low and high correlated data sets, respectively. Dashed line represents the significance level of $\alpha=0.05$.}
    \label{fig:permutation2}
\end{figure}

Figure \ref{fig:permutation1} illustrates the estimated Type-1 error for different training sample sizes for both $H_0$ case 1 and 2. In the first and second columns, we have results for the low and high correlated data sets, respectively. In $H_0$ scenarios, we expect the Type-1 error to be close to the significance level ($\alpha=0.05$). As can be seen from plots, the Type-1 error is well controlled in both cases. Figure \ref{fig:permutation2} illustrates the power of the test. In all scenarios and as expected, the power is increasing with the sample size. However, adding noise covariates slightly decreases the power when the sample size is small. As the sample size increases, the power is not affected by the presence of noise covariates.  

\subsubsection{Accuracy evaluation} 

\begin{figure}
    \centerline{\includegraphics[width=\textwidth]{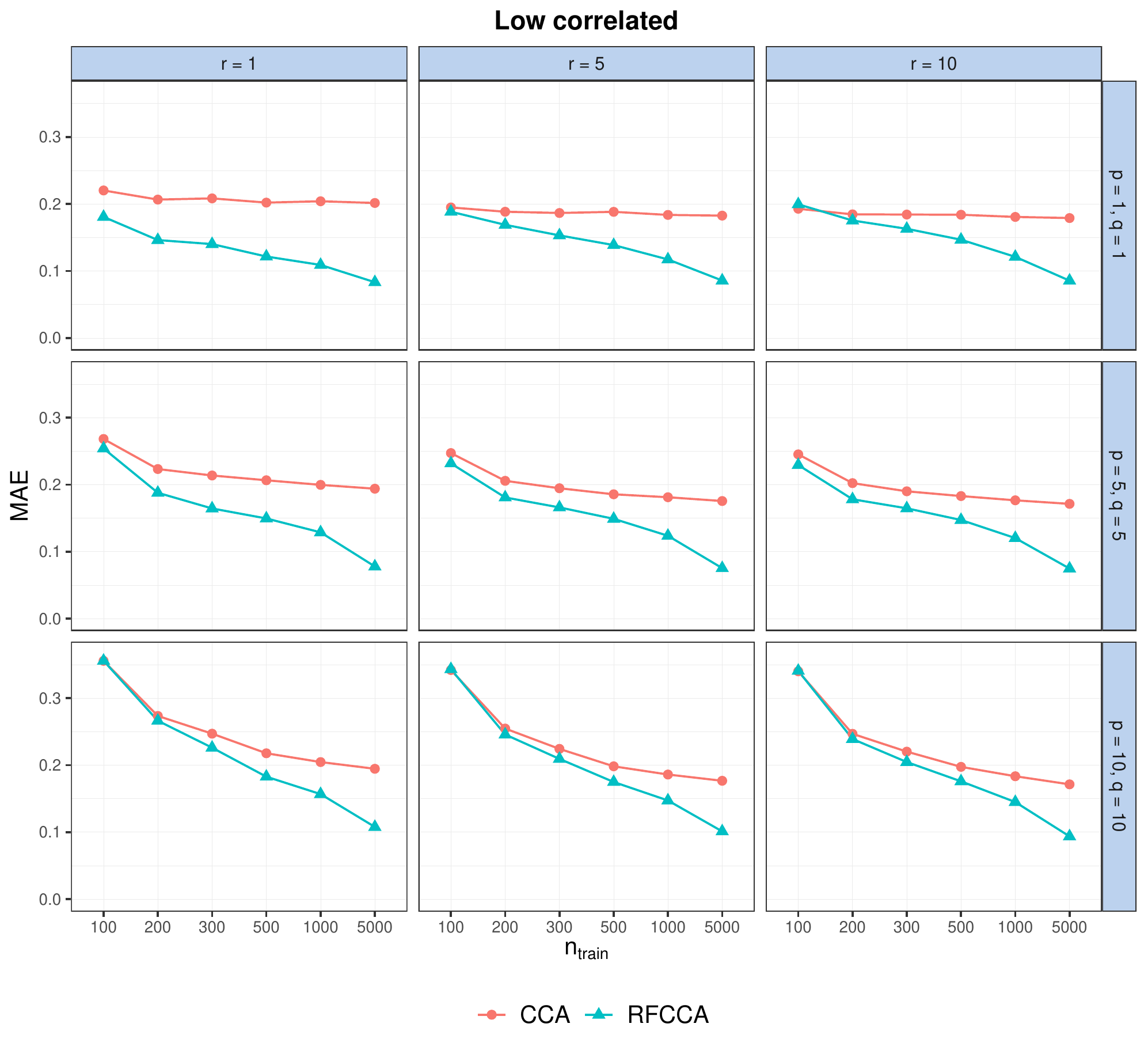}}
    \caption{Accuracy evaluation results for low correlated data sets when \texttt{nodesize} = $3\times (p+q)$. $r^{noise} = 5$ in all settings. CCA is the benchmark method. Smaller values of MAE are better.}
    \label{fig:lowcor3}
\end{figure}

\begin{figure}
    \centerline{\includegraphics[width=\textwidth]{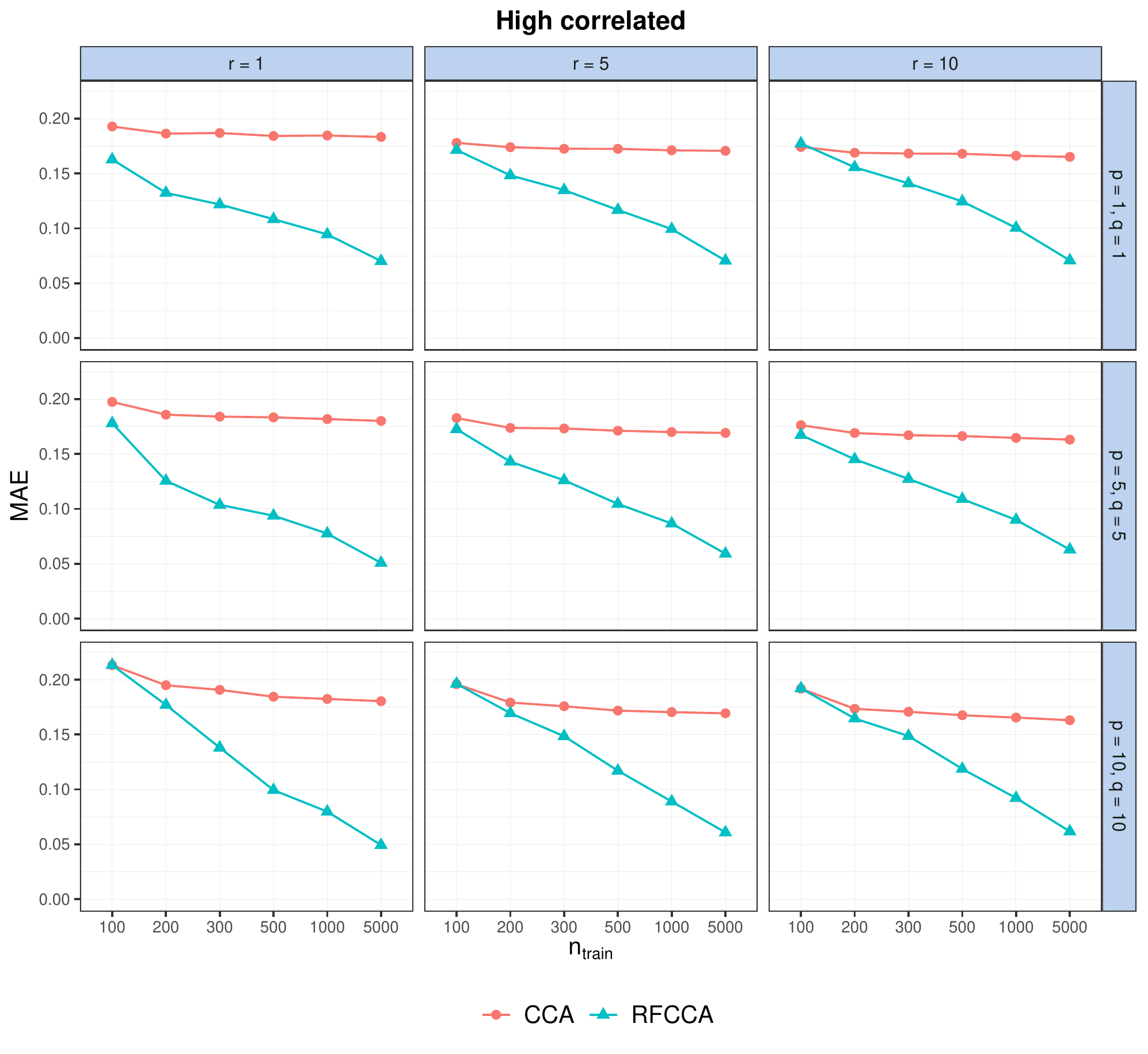}}
    \caption{Accuracy evaluation results for high correlated data sets when \texttt{nodesize} = $3\times (p+q)$. $r^{noise} = 5$ in all settings. CCA is the benchmark method. Smaller values of MAE are better.}
    \label{fig:highcor3}
\end{figure}

We provide a summary of the results in Figures \ref{fig:lowcor3} and \ref{fig:highcor3} which present the average $MAE$ over the 100 repetitions when \texttt{nodesize} = $3\times (p+q)$ for the low and high correlation settings, respectively. The plots illustrate the change in $MAE$ with increasing training sample size for different $r$ and $(p,q)$ settings. In our simulations, selecting \texttt{nodesize} as $3\times (p+q)$ leads to globally smaller or very similar $MAE$ results (see Section \ref{supp:nodesizecomparison} of the Supplementary Material for the performance comparison results for the six levels of \texttt{nodesize}). As can be seen from both figures, the proposed method and the benchmark have a similar performance, with a slight advantage for the proposed method in some cases when $n=100$. When the sample size increases, the MAE of both methods decrease but markedly faster for the proposed method. Hence, in the settings considered, a small sample size of 100 is not sufficient for the proposed method to improve over the ordinary CCA. But when the sample size increases, the proposed method successfully exploits the covariates to provide more accurate estimations of the canonical correlation, the relative gain being more important in the high correlation settings. 

In \texttt{randomForestSRC}, the default sampling for random forest training is sampling without replacement (sub-sampling), unlike the original random forest algorithm that uses bootstrapping. We investigate the effect of sampling on the performance of the proposed method on the scenarios with $n_{train}=1000$. Also, we analyze the effect of sampling on the selection of the \texttt{nodesize} parameter. The performance comparison results are presented in Section \ref{supp:samplingcomparison} of the Supplementary Material. In most of the settings, there is no significant difference in performance between sub-sampling and bootstrapping. However, in some cases, sub-sampling has slightly better accuracy than bootstrapping. Hence, we use sub-sampling in our simulations.

\subsubsection{Variable importance} \label{subsubsec:vimp}

For the variable importance, we evaluate if the estimated VIMP measures tend to rank the important variables first. In all scenarios for performance evaluation, we include noise covariates. Figures presenting the average rank, from the estimated VIMP measures, for the important variables group and noise variables group, for both low and high correlated data sets can be found in Section \ref{supp:vimpresult} of the Supplementary Material. The most important variable (the one with the highest VIMP measure) has rank 1. As ranks increases, variable importance decreases. In almost all settings, the important variables have smaller average ranks than noise variables. Only in a few settings when  $n_{train}=100$, we have  close average ranks for the important and noise variables.  

\section{Real data example} \label{sec:realdata}

Electroencephalogram (EEG) measures neuronal activity. Electrodes are distributed on the scalp to record ongoing electrical fields coming from assembles of pyramidal neurons situated in the cortex. The signal is composed of continuous variation in rhythms that can be spectrally decomposed over time through time and frequency analyses. Oscillations at different frequency bands have been found to be interdependent \citep{samiee2017time}. Only a handful of studies have assessed the effect of age on cross-frequency interdependencies. Its relationship with the intellectual level is still unknown.

In this study, 241 participants between 3 and 64 years old (113 male/128 female) and with performance IQ (pIQ) within [56, 129] and verbal IQ (vIQ) within [50, 127] were submitted to broadband noises of 50ms at 1Hz while the EEG signal was recorded using the 128 electrode EGI system (auditory evoked potentials, AEPs) (see \cite{lippe2009electrophysiological} for paradigm details). 99 participants presented a copy number variation (CNV). The data collection and preprocessing steps are described in the Supplementary Material (Section \ref{supp:eeg}). After applying the time-frequency (TF) and inter-trial coherence (ITC) analyses, we have two variables of interest, power, and phase-locking value (PLV) for each window. The time and frequency windows of interest are selected to assess low and high-frequency dependencies. In particular, the three windows for PLV are in theta waves with 3--5 Hz (100--300 ms), 3--10 Hz (4--400 ms), and 3--10 Hz (100--300 ms), and the window for power is in gamma wave with 30--50 Hz (50--150 ms). 

\begin{figure}
    \centerline{\includegraphics[width=0.7\textwidth]{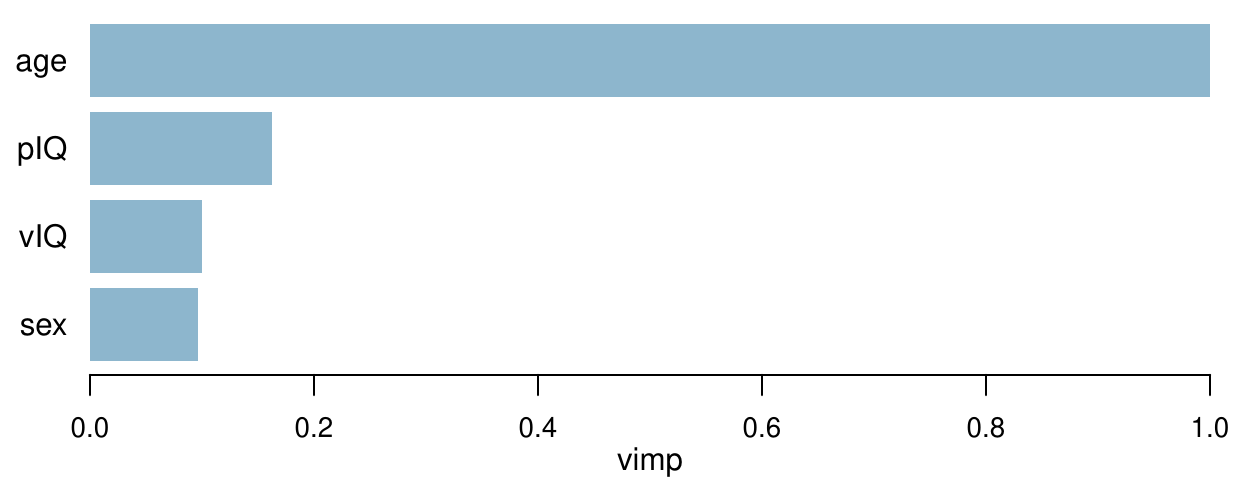}}
    \caption{Variable importance measures computed with the proposed method.}
    \label{fig:vimp}
\end{figure}

When recording auditory evoked potential, the electrodes capture information coming from the auditory cortex in the signal at the scalp level over the mid-frontal region \citep{albrecht2000development}. We want to analyze the association between the PLV variables in theta waves ($X$, $p=3$) and power in gamma wave ($Y$, $q=1$) in the mid-frontal (MF) region which is composed of Fz-Fcz in addition to four surrounding electrodes. We apply the proposed method with the subject-related covariates age, sex, pIQ, vIQ ($Z$, $r=4$) to investigate the correlation between PLV and power for the sample with $n=241$. We first perform the global significance test to evaluate the global effect of the covariates. Using 500 permutations, the estimated \textit{p}-value with \eqref{eq:pvalue} is $0.004$ and we reject the null hypothesis \eqref{eq:null}, indicating that the canonical correlation varies significantly with the covariates. Next, we apply the proposed method to the data and obtain the canonical correlation estimations. Figure \ref{fig:vimp} presents the VIMP showing that age is the most important variable followed by pIQ, vIQ and sex. Then, we use the Boruta approach \citep{boruta}, which is a permutation test based variable selection algorithm, to evaluate the statistical significance of variable importances. The main idea of this method is to compare the variable importance of the original variables with those of randomly permuted copies using statistical testing and several runs of random forests. All four covariates are selected as important variables within the significance level $\alpha = 0.01$.

\begin{figure}
    \centerline{\includegraphics[width=0.7\textwidth]{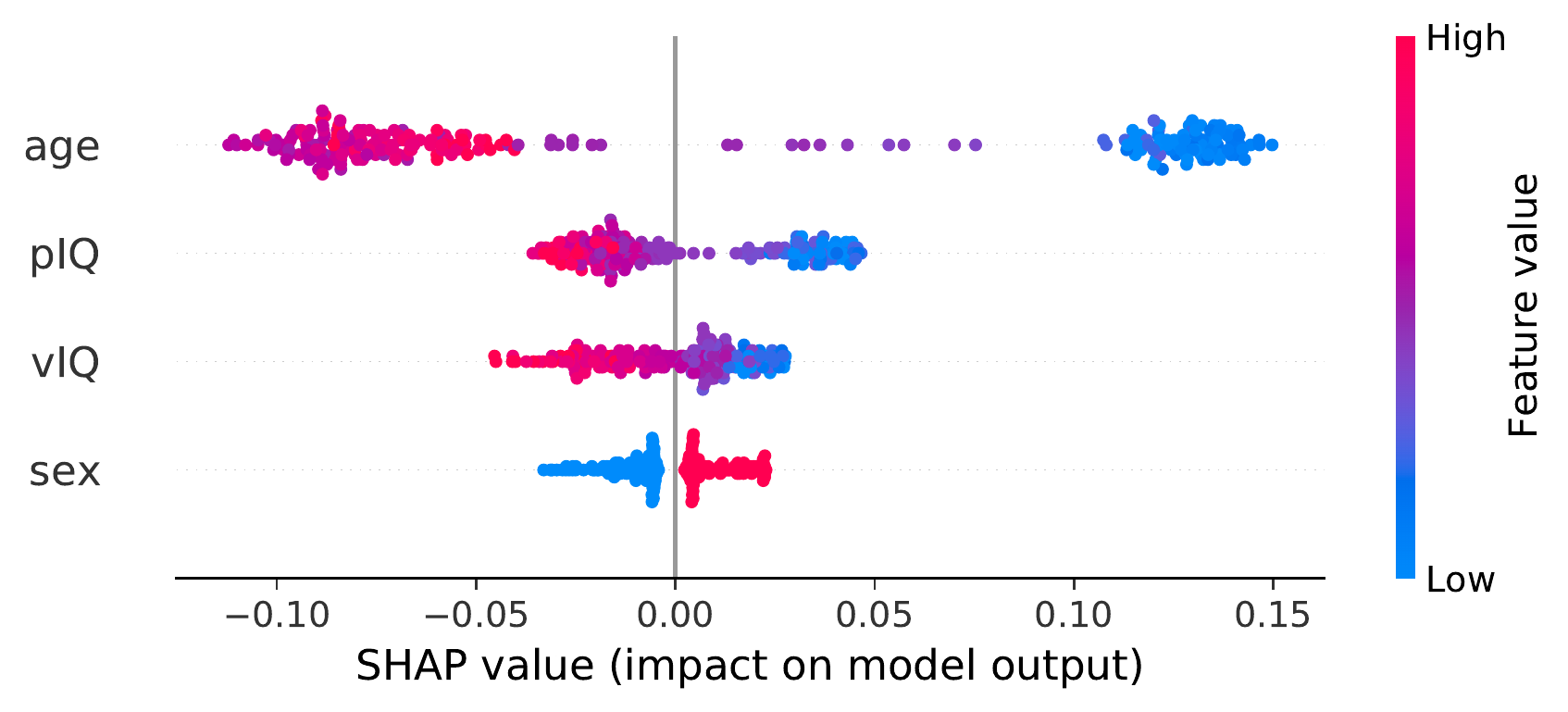}}
    \caption{SHAP summary plot.}
    \label{fig:shapsummary}
\end{figure}

We also use SHAP values \citep{lundberg2017unified} to gain additional insights. SHAP values are the contributions of each variable to the difference between the actual prediction and the expected model prediction. The sum of the contributions for each variable (SHAP values) is equal to the prediction. SHAP values show how much each variable contributes, either positively or negatively, to the individual predictions. Since the problem is unsupervised, as in the VIMP computation, we use the predicted correlations to compute the SHAP values, using \cite{lundberg2020local2global}. Figure \ref{fig:shapsummary} presents the summary plot. The covariates are ordered in the \textit{y}-axis of the plot according to their global importance showing that we obtain the same ranking as with the VIMP and thus that age is again the most important variable. The insights from this exploratory analysis fall into the expectations. Age, or brain maturation, is accompanied by important neurofunctional modifications that are reflected in the strength of the theta phase coherence and gamma power co-variations. The demonstration of a positive correlation and a strong contribution to the model in children and adolescents (3 to 20 years of age) is concordant with current brain development literature \citep{cho2015development}. Intellectual quotient, the hallmark of cognitive abilities, is found to be the second most contributive variable to the theta-gamma co-variation. Current literature focused on the theta-gamma coupling links with performances on specific cognitive tasks \citep{alekseichuk2016spatial}, and showed positive correlations. This result points to the relevance of theta-gamma co-variation in the context of abnormal neurodevelopment \citep{port2019children}.

\begin{figure}
    \centerline{\includegraphics[width=\textwidth]{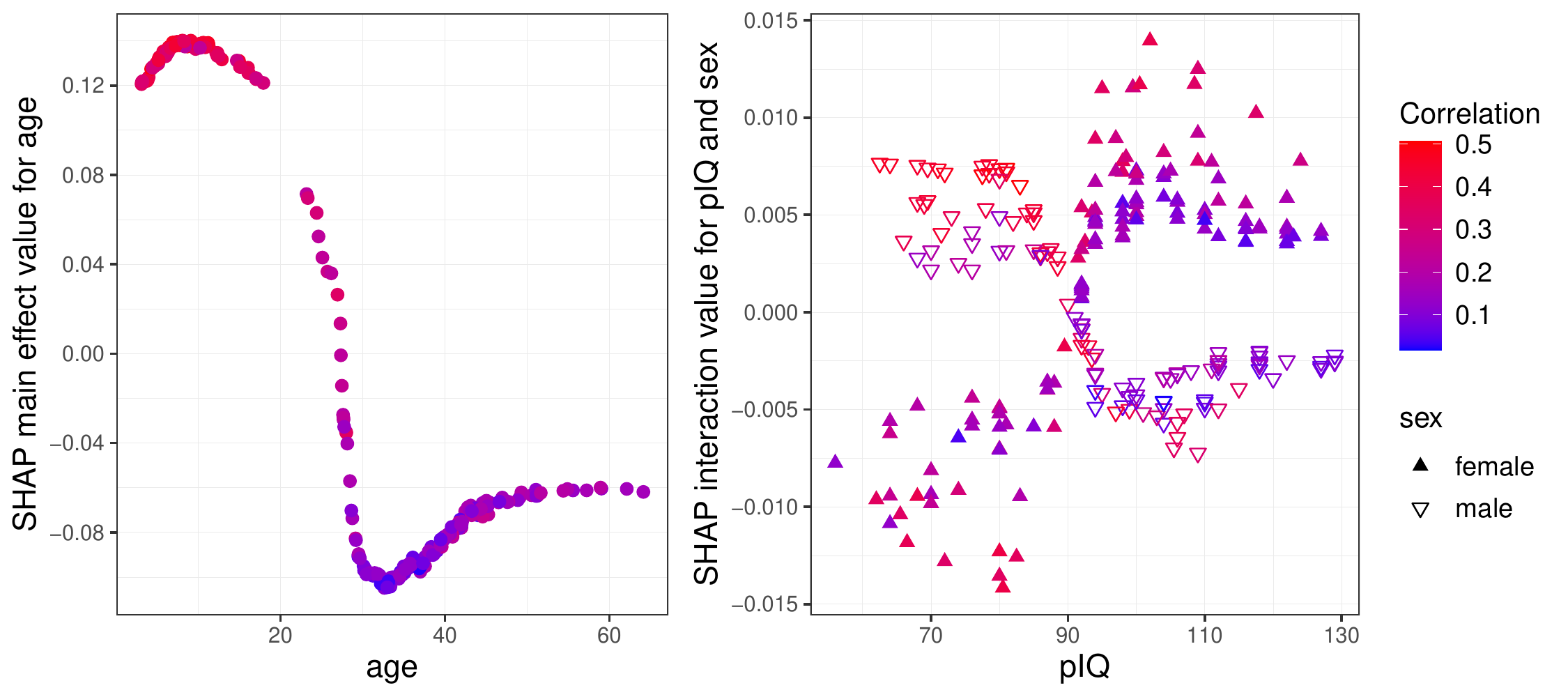}}
    \caption{(Left) SHAP main effect values for age. (Right) SHAP interaction values for sex and pIQ.}
    \label{fig:shapeffect}
\end{figure}

Left plot in Figure \ref{fig:shapeffect} shows the main effect of age on the predictions. We can see how age's attributed importance changes as its value varies. The attributed importance is on the \textit{y}-axis. The points are colored according to the predicted correlation. We can interpret this plot as the impact of age on correlation is positive and high for subjects younger than 20. Then it drops sharply reaching 0 (no impact) around 25 (note that we do not have observations around 20 and the results should be interpreted cautiously in that age range). It then continues to decrease in the negative direction, meaning the impact increases until the beginning of the 30's where it stabilizes with a slight increase afterwards (i.e. a decreasing impact). 

Right plot in Figure \ref{fig:shapeffect} presents the interaction effect between sex and pIQ (see Figure \ref{suppfig:interviq} in the Supplementary Material for the interaction effect between sex and vIQ). We see that the impact increases as we move away from the average IQ. The impact of the interaction on the theta phase coherence strength and gamma power co-variation is positive for high IQ females and negative for low IQ females. The opposite is observed in males.

As a neuroscientific conclusion, PLV in theta waves and power in gamma waves are statistically coupled and this coupling varies according to age and IQ. This confirms that the co-variation in these frequency bands relate to cognitive development. The results suggest that the co-variation between theta and gamma and intellectual capacities is nonlinear and that it follows a distinct pattern in males and females. Overall, these results indicate the importance of considering sex, intellectual capacities, and age in the study of brain signal dynamics.

\section{Concluding remarks} \label{sec:conclusion}

In this paper, we study and propose a novel random forest method to estimate the canonical correlations between two sets of variables depending on a set of subject-related covariates. The trees of the forest are built with a new splitting rule designed to form child nodes with maximum difference in the canonical correlation between the two multivariate data sets. Random forest is used to build Bag of Observations for Prediction (BOP) which can be used to compute any desired measure. We use the BOP to estimate the correlations with canonical correlation analysis for the new observations. The proposed method is flexible to various extensions. One of them is to use CCA variants such as sparse CCA \citep{witten2009penalized} and regularized CCA \citep{vinod1976canonical,leurgans1993canonical} for final canonical correlation estimation for a new observation using the constructed BOP. Another one is to build trees with alternative splitting rules such as  $n_L^{} \rho_L^2 + n_R^{} \rho_R^2$ where $\rho_L^{}$ and $\rho_R^{}$ are left and right canonical correlation estimations and $n_L^{}$ and $n_R^{}$ are left and right node sizes, respectively. It would be interesting to investigate these extensions as a future work. 

We also propose a global significance test to evaluate the global effect of the subject-related covariates and a way to compute variable importance measures. It would also be interesting to study the statistical significance of variable importance measures. The proposed method is based on the CART approach. Other tree-growing paradigms could be used, like the one that separates the variable and split point selections; the conditional inference framework \citep{hothorn2006unbiased} being one popular method. Section \ref{supp:condinf} in the Supplementary Material sketches one possible way to implement such a variant within our context. However, a limitation of the proposed method is the computational time. Computing CCA for $X \in R^{n \times p}$ and $Y \in R^{n \times q}$ has a time complexity $O\big(n(p^2+q^2)\big)$ where $n > p+q$. For each node split in each tree of the forest, we compute CCA for left and right nodes which brings a lot of CCA computations. Therefore, testing the statistical significance of variable importance of covariates at each node with a permutation test has a great computational cost. 

For simulations and real data analysis, we used the default parameter settings for \texttt{randomForestSRC} except the number of trees and the \texttt{nodesize} argument. We train random forests with 200 trees and \texttt{nodesize}$=3\times(p+q)$. The simulation study results showed that using different levels of \texttt{nodesize} could change the performance of the proposed method. In such situations, normally, we do a hyperparameter tuning to select the optimal level of the parameter. However, in our case, it is not straightforward to tune the \texttt{nodesize} parameter because we do not have an observed target. It would be interesting to find such a way to tune \texttt{nodesize} parameter as a future work. Moreover, by default, only 10 random splits are considered at each candidate splitting variable to increase the speed. Evaluating all possible splits could also improve the performance.

The proposed method can be used in other bioinformatics studies. For example, in gene-environment interaction studies \citep{caspi2006gene,hunter2005gene,ma2011varying}, the covariates ($Z$) would be the environment variables, and the two multivariate data sets ($X$ and $Y$) would correspond to brain imaging and genomic variables. In another application, the proposed method would allow us to investigate how gene expression ($Z$) can modulate the correlation between genomic ($X$) and brain imaging data ($Y$). In all these examples, the proposed algorithm can capture nonlinear interactions, which is new compared to existing approaches.

\section*{Acknowledgements}

We would like to thank two anonymous reviewers for  their interesting and constructive comments that lead to an improved version of this article. \vspace*{-12pt}

\section*{Funding}

This research was supported by the Natural Sciences and Engineering Research Council of Canada (NSERC), and by Fondation HEC Montr\'eal. \vspace*{-12pt}

\newpage
\bibliography{refs}

\begin{thebibliography}{}

\bibitem[Akaho(2001)Akaho]{Akaho01akernel}
Akaho, S. (2001).
\newblock A kernel method for canonical correlation analysis.
\newblock In {\em In Proceedings of the International Meeting of the
  Psychometric Society (IMPS2001\/}. Springer-Verlag.

\bibitem[Albrecht {\em et~al.}(2000)Albrecht, Suchodoletz, and
  Uwer]{albrecht2000development}
Albrecht, R.  {\em et~al.} (2000).
\newblock The development of auditory evoked dipole source activity from
  childhood to adulthood.
\newblock {\em Clinical Neurophysiology\/}, {\bf 111}(12), 2268--2276.

\bibitem[Alekseichuk {\em et~al.}(2016)Alekseichuk, Turi, de~Lara, Antal, and
  Paulus]{alekseichuk2016spatial}
Alekseichuk, I.  {\em et~al.} (2016).
\newblock Spatial working memory in humans depends on theta and high gamma
  synchronization in the prefrontal cortex.
\newblock {\em Current Biology\/}, {\bf 26}(12), 1513--1521.

\bibitem[Andrew {\em et~al.}(2013)Andrew, Arora, Bilmes, and
  Livescu]{andrew2013deep}
Andrew, G.  {\em et~al.} (2013).
\newblock Deep canonical correlation analysis.
\newblock In {\em International conference on machine learning\/}, pages
  1247--1255.

\bibitem[Athey {\em et~al.}(2019)Athey, Tibshirani, Wager, {\em
  et~al.}]{athey2019generalized}
Athey, S.  {\em et~al.} (2019).
\newblock Generalized random forests.
\newblock {\em The Annals of Statistics\/}, {\bf 47}(2), 1148--1178.

\bibitem[Bach and Jordan(2002)Bach and Jordan]{bach2002kernel}
Bach, F.~R. and Jordan, M.~I. (2002).
\newblock Kernel independent component analysis.
\newblock {\em Journal of machine learning research\/}, {\bf 3}(Jul), 1--48.

\bibitem[Branco {\em et~al.}(2005)Branco, Croux, Filzmoser, and
  Oliveira]{branco2005robust}
Branco, J.~A.  {\em et~al.} (2005).
\newblock Robust canonical correlations: A comparative study.
\newblock {\em Computational Statistics\/}, {\bf 20}(2), 203--229.

\bibitem[Breiman(2001)Breiman]{breiman2001random}
Breiman, L. (2001).
\newblock Random forests.
\newblock {\em Machine learning\/}, {\bf 45}(1), 5--32.

\bibitem[Breiman {\em et~al.}(1984)Breiman, Friedman, Stone, and
  Olshen]{breiman1984classification}
Breiman, L.  {\em et~al.} (1984).
\newblock {\em Classification and regression trees\/}.
\newblock CRC press.

\bibitem[{Cancer Genome Atlas Network}(2012){Cancer Genome Atlas
  Network}]{cancer2012comprehensive}
{Cancer Genome Atlas Network} (2012).
\newblock Comprehensive molecular portraits of human breast tumours.
\newblock {\em Nature\/}, {\bf 490}(7418), 61.

\bibitem[Caspi and Moffitt(2006)Caspi and Moffitt]{caspi2006gene}
Caspi, A. and Moffitt, T.~E. (2006).
\newblock Gene--environment interactions in psychiatry: joining forces with
  neuroscience.
\newblock {\em Nature Reviews Neuroscience\/}, {\bf 7}(7), 583--590.

\bibitem[Cho {\em et~al.}(2015)Cho, Walker, Polizzotto, Wozny, Fissell, Chen,
  and Lewis]{cho2015development}
Cho, R.~Y.  {\em et~al.} (2015).
\newblock Development of sensory gamma oscillations and cross-frequency
  coupling from childhood to early adulthood.
\newblock {\em Cerebral cortex\/}, {\bf 25}(6), 1509--1518.

\bibitem[Choi {\em et~al.}(2020)Choi, Li, Liu, and Zeng]{choi2020recursive}
Choi, D.  {\em et~al.} (2020).
\newblock A recursive partitioning approach for subgroup identification in
  brain--behaviour correlation analysis.
\newblock {\em Pattern Analysis and Applications\/}, {\bf 23}(1), 161--177.

\bibitem[Cruz-Cano and Lee(2014)Cruz-Cano and Lee]{cruz2014fast}
Cruz-Cano, R. and Lee, M.-L.~T. (2014).
\newblock Fast regularized canonical correlation analysis.
\newblock {\em Computational Statistics \& Data Analysis\/}, {\bf 70}, 88--100.

\bibitem[Davis {\em et~al.}(2008)Davis, Dennis, Daselaar, Fleck, and
  Cabeza]{davis2008pasa}
Davis, S.~W.  {\em et~al.} (2008).
\newblock Que pasa? the posterior--anterior shift in aging.
\newblock {\em Cerebral cortex\/}, {\bf 18}(5), 1201--1209.

\bibitem[{ENCODE Project Consortium}(2012){ENCODE Project
  Consortium}]{encode2012integrated}
{ENCODE Project Consortium} (2012).
\newblock An integrated encyclopedia of dna elements in the human genome.
\newblock {\em Nature\/}, {\bf 489}(7414), 57--74.

\bibitem[Ewerbring {\em et~al.}(1990)Ewerbring {\em
  et~al.}]{ewerbring1990canonical}
Ewerbring, L.~M. {\em et~al.} (1990).
\newblock Canonical correlations and generalized svd: applications and new
  algorithms.
\newblock In {\em Advances in Parallel Computing\/}, volume~1, pages 37--52.
  Elsevier.

\bibitem[Fratello {\em et~al.}(2017)Fratello, Caiazzo, Trojsi, Russo, Tedeschi,
  Tagliaferri, and Esposito]{fratello2017multi}
Fratello, M.  {\em et~al.} (2017).
\newblock Multi-view ensemble classification of brain connectivity images for
  neurodegeneration type discrimination.
\newblock {\em Neuroinformatics\/}, {\bf 15}(2), 199--213.

\bibitem[Hanna {\em et~al.}(2010)Hanna, Molfenter, Cliffe, Chau, and
  Steele]{hanna2010anthropometric}
Hanna, F.  {\em et~al.} (2010).
\newblock Anthropometric and demographic correlates of dual-axis swallowing
  accelerometry signal characteristics: a canonical correlation analysis.
\newblock {\em Dysphagia\/}, {\bf 25}(2), 94--103.

\bibitem[Hardoon and Shawe-Taylor(2011)Hardoon and
  Shawe-Taylor]{hardoon2011sparse}
Hardoon, D.~R. and Shawe-Taylor, J. (2011).
\newblock Sparse canonical correlation analysis.
\newblock {\em Machine Learning\/}, {\bf 83}(3), 331--353.

\bibitem[Hardoon {\em et~al.}(2004)Hardoon, Szedmak, and
  Shawe-Taylor]{hardoon2004canonical}
Hardoon, D.~R.  {\em et~al.} (2004).
\newblock Canonical correlation analysis: An overview with application to
  learning methods.
\newblock {\em Neural computation\/}, {\bf 16}(12), 2639--2664.

\bibitem[Healy(1957)Healy]{healy1957rotation}
Healy, M. (1957).
\newblock A rotation method for computing canonical correlations.
\newblock {\em Mathematics of Computation\/}, {\bf 11}(58), 83--86.

\bibitem[Hotelling(1936)Hotelling]{hotelling1936}
Hotelling, H. (1936).
\newblock Relations between two sets of variates.
\newblock {\em Biometrika\/}, {\bf 28}(3/4), 321--377.

\bibitem[Hothorn {\em et~al.}(2004)Hothorn, Lausen, Benner, and
  Radespiel-Tr{\"o}ger]{hothorn2004bagging}
Hothorn, T.  {\em et~al.} (2004).
\newblock Bagging survival trees.
\newblock {\em Statistics in medicine\/}, {\bf 23}(1), 77--91.

\bibitem[Hothorn {\em et~al.}(2006)Hothorn, Hornik, and
  Zeileis]{hothorn2006unbiased}
Hothorn, T.  {\em et~al.} (2006).
\newblock Unbiased recursive partitioning: A conditional inference framework.
\newblock {\em Journal of Computational and Graphical statistics\/}, {\bf
  15}(3), 651--674.

\bibitem[Hunter(2005)Hunter]{hunter2005gene}
Hunter, D.~J. (2005).
\newblock Gene--environment interactions in human diseases.
\newblock {\em Nature Reviews Genetics\/}, {\bf 6}(4), 287--298.

\bibitem[Ishwaran and Kogalur(2020)Ishwaran and Kogalur]{rfsrc}
Ishwaran, H. and Kogalur, U. (2020).
\newblock {\em Fast Unified Random Forests for Survival, Regression, and
  Classification (RF-SRC)\/}.
\newblock R package version 2.9.3.

\bibitem[Kettenring(1971)Kettenring]{kettenring1971canonical}
Kettenring, J.~R. (1971).
\newblock Canonical analysis of several sets of variables.
\newblock {\em Biometrika\/}, {\bf 58}(3), 433--451.

\bibitem[Kursa and Rudnicki(2010)Kursa and Rudnicki]{boruta}
Kursa, M.~B. and Rudnicki, W.~R. (2010).
\newblock Feature selection with the {Boruta} package.
\newblock {\em Journal of Statistical Software\/}, {\bf 36}(11), 1--13.

\bibitem[Leurgans {\em et~al.}(1993)Leurgans, Moyeed, and
  Silverman]{leurgans1993canonical}
Leurgans, S.~E.  {\em et~al.} (1993).
\newblock Canonical correlation analysis when the data are curves.
\newblock {\em Journal of the Royal Statistical Society: Series B
  (Methodological)\/}, {\bf 55}(3), 725--740.

\bibitem[Li and Jung(2017)Li and Jung]{li2017incorporating}
Li, G. and Jung, S. (2017).
\newblock Incorporating covariates into integrated factor analysis of
  multi-view data.
\newblock {\em Biometrics\/}, {\bf 73}(4), 1433--1442.

\bibitem[Li {\em et~al.}(2010)Li, Luo, and Gong]{li2010gender}
Li, T.  {\em et~al.} (2010).
\newblock Gender-specific hemodynamics in prefrontal cortex during a verbal
  working memory task by near-infrared spectroscopy.
\newblock {\em Behavioural brain research\/}, {\bf 209}(1), 148--153.

\bibitem[Li {\em et~al.}(2018)Li, Wu, and Ngom]{li2018review}
Li, Y.  {\em et~al.} (2018).
\newblock A review on machine learning principles for multi-view biological
  data integration.
\newblock {\em Briefings in bioinformatics\/}, {\bf 19}(2), 325--340.

\bibitem[Lin and Jeon(2006)Lin and Jeon]{lin2006random}
Lin, Y. and Jeon, Y. (2006).
\newblock Random forests and adaptive nearest neighbors.
\newblock {\em Journal of the American Statistical Association\/}, {\bf
  101}(474), 578--590.

\bibitem[Lipp{\'e} {\em et~al.}(2009)Lipp{\'e}, Martinez-Montes, Arcand, and
  Lassonde]{lippe2009electrophysiological}
Lipp{\'e}, S.  {\em et~al.} (2009).
\newblock Electrophysiological study of auditory development.
\newblock {\em Neuroscience\/}, {\bf 164}(3), 1108--1118.

\bibitem[Lundberg and Lee(2017)Lundberg and Lee]{lundberg2017unified}
Lundberg, S.~M. and Lee, S.-I. (2017).
\newblock A unified approach to interpreting model predictions.
\newblock In {\em Advances in neural information processing systems\/}, pages
  4765--4774.

\bibitem[Lundberg {\em et~al.}(2020)Lundberg, Erion, Chen, DeGrave, Prutkin,
  Nair, Katz, Himmelfarb, Bansal, and Lee]{lundberg2020local2global}
Lundberg, S.~M.  {\em et~al.} (2020).
\newblock From local explanations to global understanding with explainable ai
  for trees.
\newblock {\em Nature Machine Intelligence\/}, {\bf 2}(1), 2522--5839.

\bibitem[Ma {\em et~al.}(2011)Ma, Yang, Romero, and Cui]{ma2011varying}
Ma, S.  {\em et~al.} (2011).
\newblock Varying coefficient model for gene--environment interaction: a
  non-linear look.
\newblock {\em Bioinformatics\/}, {\bf 27}(15), 2119--2126.

\bibitem[Melzer {\em et~al.}(2001)Melzer, Reiter, and
  Bischof]{melzer2001nonlinear}
Melzer, T.  {\em et~al.} (2001).
\newblock Nonlinear feature extraction using generalized canonical correlation
  analysis.
\newblock In {\em International Conference on Artificial Neural Networks\/},
  pages 353--360. Springer.

\bibitem[Meng {\em et~al.}(2016)Meng, Zeleznik, Thallinger, Kuster, Gholami,
  and Culhane]{meng2016dimension}
Meng, C.  {\em et~al.} (2016).
\newblock Dimension reduction techniques for the integrative analysis of
  multi-omics data.
\newblock {\em Briefings in bioinformatics\/}, {\bf 17}(4), 628--641.

\bibitem[Michaeli {\em et~al.}(2016)Michaeli, Wang, and
  Livescu]{michaeli2016nonparametric}
Michaeli, T.  {\em et~al.} (2016).
\newblock Nonparametric canonical correlation analysis.
\newblock In {\em International Conference on Machine Learning\/}, pages
  1967--1976.

\bibitem[Mihalik {\em et~al.}(2020)Mihalik, Ferreira, Moutoussis, Ziegler,
  Adams, Rosa, Prabhu, de~Oliveira, Pereira, Bullmore, {\em
  et~al.}]{mihalik2020multiple}
Mihalik, A.  {\em et~al.} (2020).
\newblock Multiple holdouts with stability: Improving the generalizability of
  machine learning analyses of brain--behavior relationships.
\newblock {\em Biological psychiatry\/}, {\bf 87}(4), 368--376.

\bibitem[Min {\em et~al.}(2017)Min, Lee, and Yoon]{min2017deep}
Min, S.  {\em et~al.} (2017).
\newblock Deep learning in bioinformatics.
\newblock {\em Briefings in bioinformatics\/}, {\bf 18}(5), 851--869.

\bibitem[Moradian {\em et~al.}(2017)Moradian, Larocque, and
  Bellavance]{moradian2017l1}
Moradian, H.  {\em et~al.} (2017).
\newblock L1 splitting rules in survival forests.
\newblock {\em Lifetime data analysis\/}, {\bf 23}(4), 671.

\bibitem[Moradian {\em et~al.}(2019)Moradian, Larocque, and
  Bellavance]{moradian2019survival}
Moradian, H.  {\em et~al.} (2019).
\newblock Survival forests for data with dependent censoring.
\newblock {\em Statistical methods in medical research\/}, {\bf 28}(2),
  445--461.

\bibitem[Moser {\em et~al.}(2018)Moser, Doucet, Lee, Rasgon, Krinsky, Leibu,
  Ing, Schumann, Rasgon, and Frangou]{moser2018multivariate}
Moser, D.~A.  {\em et~al.} (2018).
\newblock Multivariate associations among behavioral, clinical, and multimodal
  imaging phenotypes in patients with psychosis.
\newblock {\em JAMA psychiatry\/}, {\bf 75}(4), 386--395.

\bibitem[Pezeshki {\em et~al.}(2004)Pezeshki, Scharf, Azimi-Sadjadi, and
  Lundberg]{pezeshki2004empirical}
Pezeshki, A.  {\em et~al.} (2004).
\newblock Empirical canonical correlation analysis in subspaces.
\newblock In {\em Conference Record of the Thirty-Eighth Asilomar Conference on
  Signals, Systems and Computers, 2004.}, volume~1, pages 994--997. IEEE.

\bibitem[Port {\em et~al.}(2019)Port, Dipiero, Ku, Liu, Blaskey, Kuschner,
  Edgar, Roberts, and Berman]{port2019children}
Port, R.~G.  {\em et~al.} (2019).
\newblock Children with autism spectrum disorder demonstrate regionally
  specific altered resting-state phase--amplitude coupling.
\newblock {\em Brain Connectivity\/}, {\bf 9}(5), 425--436.

\bibitem[Roy and Larocque(2020)Roy and Larocque]{roy2020prediction}
Roy, M.-H. and Larocque, D. (2020).
\newblock Prediction intervals with random forests.
\newblock {\em Statistical Methods in Medical Research\/}, {\bf 29}(1),
  205--229.

\bibitem[Samiee and Baillet(2017)Samiee and Baillet]{samiee2017time}
Samiee, S. and Baillet, S. (2017).
\newblock Time-resolved phase-amplitude coupling in neural oscillations.
\newblock {\em NeuroImage\/}, {\bf 159}, 270--279.

\bibitem[Sun(2013)Sun]{sun2013survey}
Sun, S. (2013).
\newblock A survey of multi-view machine learning.
\newblock {\em Neural computing and applications\/}, {\bf 23}(7-8), 2031--2038.

\bibitem[Tabib and Larocque(2020)Tabib and Larocque]{tabib2020non}
Tabib, S. and Larocque, D. (2020).
\newblock Non-parametric individual treatment effect estimation for survival
  data with random forests.
\newblock {\em Bioinformatics\/}, {\bf 36}(2), 629--636.

\bibitem[Vinod(1976)Vinod]{vinod1976canonical}
Vinod, H.~D. (1976).
\newblock Canonical ridge and econometrics of joint production.
\newblock {\em Journal of econometrics\/}, {\bf 4}(2), 147--166.

\bibitem[Wilms and Croux(2015)Wilms and Croux]{wilms2015sparse}
Wilms, I. and Croux, C. (2015).
\newblock Sparse canonical correlation analysis from a predictive point of
  view.
\newblock {\em Biometrical Journal\/}, {\bf 57}(5), 834--851.

\bibitem[Witten {\em et~al.}(2009)Witten, Tibshirani, and
  Hastie]{witten2009penalized}
Witten, D.~M.  {\em et~al.} (2009).
\newblock A penalized matrix decomposition, with applications to sparse
  principal components and canonical correlation analysis.
\newblock {\em Biostatistics\/}, {\bf 10}(3), 515--534.

\end{thebibliography}


\begin{thebibliography}{}

\bibitem[Bach and Jordan(2002)Bach and Jordan]{bach2002kernel}
Bach, F.~R. and Jordan, M.~I. (2002).
\newblock Kernel independent component analysis.
\newblock {\em Journal of machine learning research\/}, {\bf 3}(Jul), 1--48.

\bibitem[Bj{\"o}rck and Golub(1973)Bj{\"o}rck and Golub]{bjorck1973numerical}
Bj{\"o}rck, {\AA}. and Golub, G.~H. (1973).
\newblock Numerical methods for computing angles between linear subspaces.
\newblock {\em Mathematics of computation\/}, {\bf 27}(123), 579--594.

\bibitem[Branco {\em et~al.}(2005)Branco, Croux, Filzmoser, and
  Oliveira]{branco2005robust}
Branco, J.~A.  {\em et~al.} (2005).
\newblock Robust canonical correlations: A comparative study.
\newblock {\em Computational Statistics\/}, {\bf 20}(2), 203--229.

\bibitem[Breiman(2001)Breiman]{breiman2001random}
Breiman, L. (2001).
\newblock Random forests.
\newblock {\em Machine learning\/}, {\bf 45}(1), 5--32.

\bibitem[Delorme and Makeig(2004)Delorme and Makeig]{delorme2004eeglab}
Delorme, A. and Makeig, S. (2004).
\newblock Eeglab: an open source toolbox for analysis of single-trial eeg
  dynamics including independent component analysis.
\newblock {\em Journal of neuroscience methods\/}, {\bf 134}(1), 9--21.

\bibitem[Delorme {\em et~al.}(2007)Delorme, Sejnowski, and
  Makeig]{delorme2007enhanced}
Delorme, A.  {\em et~al.} (2007).
\newblock Enhanced detection of artifacts in eeg data using higher-order
  statistics and independent component analysis.
\newblock {\em Neuroimage\/}, {\bf 34}(4), 1443--1449.

\bibitem[Do~Q(2012)Do~Q]{do2012numerically}
Do~Q, L. (2012).
\newblock Numerically efficient methods for solving least squares problems.

\bibitem[Ewerbring {\em et~al.}(1990)Ewerbring {\em
  et~al.}]{ewerbring1990canonical}
Ewerbring, L.~M. {\em et~al.} (1990).
\newblock Canonical correlations and generalized svd: applications and new
  algorithms.
\newblock In {\em Advances in Parallel Computing\/}, volume~1, pages 37--52.
  Elsevier.

\bibitem[Golub(1969)Golub]{golub1969matrix}
Golub, G.~H. (1969).
\newblock Matrix decompositions and statistical calculations.
\newblock In {\em Statistical computation\/}, pages 365--397. Elsevier.

\bibitem[Hardoon {\em et~al.}(2004)Hardoon, Szedmak, and
  Shawe-Taylor]{hardoon2004canonical}
Hardoon, D.~R.  {\em et~al.} (2004).
\newblock Canonical correlation analysis: An overview with application to
  learning methods.
\newblock {\em Neural computation\/}, {\bf 16}(12), 2639--2664.

\bibitem[Healy(1957)Healy]{healy1957rotation}
Healy, M. (1957).
\newblock A rotation method for computing canonical correlations.
\newblock {\em Mathematics of Computation\/}, {\bf 11}(58), 83--86.

\bibitem[Herrmann {\em et~al.}(2005)Herrmann, Grigutsch, and
  Busch]{herrmann200511}
Herrmann, C.~S.  {\em et~al.} (2005).
\newblock Eeg oscillations and wavelet analysis.
\newblock {\em Event-related potentials: A methods handbook\/}, pages 229--259.

\bibitem[Hotelling(1936)Hotelling]{hotelling1936}
Hotelling, H. (1936).
\newblock Relations between two sets of variates.
\newblock {\em Biometrika\/}, {\bf 28}(3/4), 321--377.

\bibitem[Hothorn {\em et~al.}(2006)Hothorn, Hornik, and
  Zeileis]{hothorn2006unbiased}
Hothorn, T.  {\em et~al.} (2006).
\newblock Unbiased recursive partitioning: A conditional inference framework.
\newblock {\em Journal of Computational and Graphical statistics\/}, {\bf
  15}(3), 651--674.

\bibitem[Ishwaran(2007)Ishwaran]{ishwaran2007variable}
Ishwaran, H. (2007).
\newblock Variable importance in binary regression trees and forests.
\newblock {\em Electronic Journal of Statistics\/}, {\bf 1}, 519--537.

\bibitem[Lachaux {\em et~al.}(1999)Lachaux, Rodriguez, Martinerie, and
  Varela]{lachaux1999measuring}
Lachaux, J.-P.  {\em et~al.} (1999).
\newblock Measuring phase synchrony in brain signals.
\newblock {\em Human brain mapping\/}, {\bf 8}(4), 194--208.

\bibitem[Makeig {\em et~al.}(2004)Makeig, Debener, Onton, and
  Delorme]{makeig2004mining}
Makeig, S.  {\em et~al.} (2004).
\newblock Mining event-related brain dynamics.
\newblock {\em Trends in cognitive sciences\/}, {\bf 8}(5), 204--210.

\bibitem[Tallon-Baudry and Bertrand(1999)Tallon-Baudry and
  Bertrand]{tallon1999oscillatory}
Tallon-Baudry, C. and Bertrand, O. (1999).
\newblock Oscillatory gamma activity in humans and its role in object
  representation.
\newblock {\em Trends in cognitive sciences\/}, {\bf 3}(4), 151--162.

\bibitem[Tucker(1993)Tucker]{tucker1993spatial}
Tucker, D.~M. (1993).
\newblock Spatial sampling of head electrical fields: the geodesic sensor net.
\newblock {\em Electroencephalography and clinical neurophysiology\/}, {\bf
  87}(3), 154--163.

\bibitem[Wilms and Croux(2015)Wilms and Croux]{wilms2015sparse}
Wilms, I. and Croux, C. (2015).
\newblock Sparse canonical correlation analysis from a predictive point of
  view.
\newblock {\em Biometrical Journal\/}, {\bf 57}(5), 834--851.

\end{thebibliography}

\newpage
\appendix
\renewcommand{\thesection}{\arabic{section}}
\renewcommand{\thefigure}{A\arabic{figure}}
\renewcommand{\theHfigure}{A\arabic{figure}}

\begin{center}
\LARGE{\textbf{Supplementary Material for \\Conditional canonical correlation estimation based on covariates with random forests}}
\end{center}
\vspace{0.5cm}

\renewcommand{\thefigure}{\arabic{figure}}

\setcounter{figure}{0}

\section{Motivating examples}

\subsection{Example with univariate $X$ and $Y$}

\begin{figure}
    \centerline{\includegraphics[width=0.7\textwidth]{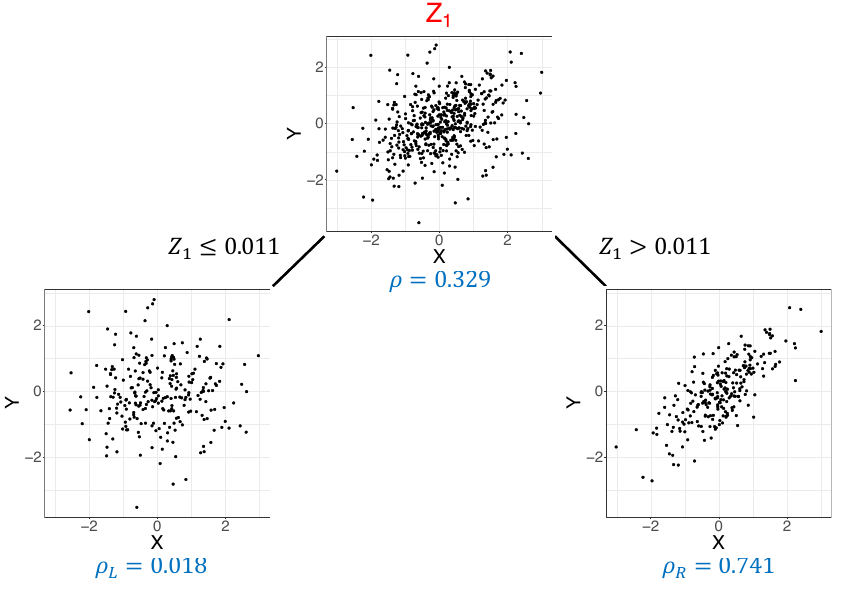}}
    \caption{Illustration for the example, single split of a single decision tree.}
    \label{fig:toyextreesup}
\end{figure}

We generated a data set with $Z=(Z_1,\ldots,Z_{10}) \in R^{n \times 10}$, $X \in R^{n \times 1}$ and $Y \in R^{n \times 1}$ where $n = 500$, $Z_i \sim N(0,1) \ \forall i = \{1,2,...,10\}$ and $(X,Y) \sim MVN(0,\Sigma)$ where $\Sigma = \left( \begin{smallmatrix} 
	1 & \rho\\
	\rho & 1
\end{smallmatrix} \right)$. The correlation between $X$ and $Y$ is a function of $Z_1$ given by 
\begin{align*}
\rho = \left\{ \begin{array}{lc} 
    0 & \hspace{5mm} Z_1 \leq 0 \\
    0.8 & \hspace{5mm} Z_1 > 0 \\
    \end{array} 
\right.
\end{align*}
where $\rho$ is the population correlation between $X$ and $Y$. The sample correlation between $X$ and $Y$, as a function of $Z_1$, is 
\begin{align*}
r(X,Y) = \left\{ \begin{array}{lc} 
    0.017 & \hspace{5mm} Z_1 \leq 0 \\
    0.737 & \hspace{5mm} Z_1 > 0 \\
    \end{array} 
\right.
\end{align*}
whereas the sample correlation between all $X$ and $Y$ is 0.329. Our aim is to identify the two groups of observations having the most different correlation. After applying the proposed method with a single tree and a single split, we have two groups of observations. Figure \ref{fig:toyextreesup} illustrates the single split of the method. The observations are grouped according to their $Z_1$ values, the first group has $Z_1 \leq 0.011$ and the second group has $Z_1 > 0.011$. The correlation between $X$ and $Y$ in those two groups are respectively 0.018 and 0.741. Hence, the proposed method was able to identify the two groups. 

\subsection{Example with multivariate $X$ and $Y$}

As a second example, now with multivariate $X$ and $Y$, we have a data set with  $Z \in R^{n \times 10}$, $X \in R^{n \times 2}$ and $Y \in R^{n \times 2}$ where $n = 500$, $Z_i \sim N(0,1) \ \forall i = \{1,2,...,10\}$. The canonical correlation between $X$ and $Y$ is a function of $Z_1$ 
\begin{align*}
\rho_{CCA} = \left\{ \begin{array}{lc} 
    0.2 & \hspace{5mm} Z_1 \leq 0 \\
    0.8 & \hspace{5mm} Z_1 > 0 \\
    \end{array} 
\right.
\end{align*}
where $\rho_{CCA}$ is the population CCA correlation between $X$ and $Y$. The sample CCA correlations between $X$ and $Y$ as a function of $Z_1$ is
\begin{align*}
r_{CCA}(X,Y) = \left\{ \begin{array}{lc} 
    0.237 & \hspace{5mm} Z_1 \leq 0 \\
    0.785 & \hspace{5mm} Z_1 > 0 \\
    \end{array} 
\right.
\end{align*}
whereas the sample CCA correlation between all $X$ and $Y$ is 0.535. Again, after applying the proposed method with a single tree and single split, we have two groups of observations. The observations are grouped according to their $Z_1$ values, the first group has $Z_1 \leq 0.014$ and the second group has $Z_1 > 0.014$. The CCA correlation between $X$ and $Y$ in those two groups are respectively 0.237 and 0.785. Thus, again, the proposed method was able to identify the two groups.

\section{Computing canonical correlations}

Canonical correlation analysis, firstly introduced in \citesm{hotelling1936}, seeks vectors of $a \in R^{p}$ and $b \in R^{q}$ for  two multivariate data sets $X \in R^{n \times p}$ and $Y \in R^{n \times q}$ such that $Xa$ and $Yb$ are maximally linearly correlated. We can formulate the problem 
\begin{equation} \label{maxcca1}
    (a^*, b^*) = \operatorname*{argmax}_{a,b} corr(Xa, Yb)
\end{equation}
where
\begin{equation*}
    corr(Xa, Yb) = \frac{a^T \Sigma_{XY} b}{\sqrt{a^T \Sigma_{XX} a} \sqrt{b^T \Sigma_{YY} b}},
\end{equation*}

and where $\Sigma_{XX}$ and $\Sigma_{YY}$ are the covariance matrices of $X$ and $Y$, respectively, and $\Sigma_{XY}$ is the cross-covariance matrix. The choice of rescaling of $a$ and $b$ does not affect the $corr(Xa, Yb)$, so we can add the constraints $a^T \Sigma_{XX} a = 1$, $b^T \Sigma_{YY} b = 1$ to the maximization problem \eqref{maxcca1}.

In \citesm{hotelling1936}, the CCA solution is based on two similar-looking equations 
\begin{equation} \label{ccaeq1}
    (\Sigma_{YX} \Sigma_{XX}^{-1} \Sigma_{XY} - \rho^2 \Sigma_{YY}) y = 0
\end{equation}
\begin{equation} \label{ccaeq2}
    (\Sigma_{XY} \Sigma_{YY}^{-1} \Sigma_{YX} - \rho^2 \Sigma_{XX}) x = 0
\end{equation}
where $\Sigma_{YX}$ is the transpose of $\Sigma_{XY}$. We can find canonical correlations and $(a^*,b^*)$ by solving two standard eigenvalue problems. Basically, eigenvalues of \eqref{ccaeq1} and \eqref{ccaeq2} are the same and equal to the squared canonical correlations. 

In addition to solving standard eigenvalue problem, there are some alternative ways to solve CCA. We can solve the generalized eigenvalue problem \citepsm{hardoon2004canonical, bach2002kernel}. We can find the solution of CCA by alternating least squares regression  \citepsm{branco2005robust,wilms2015sparse}. Alternatively, we can use singular value decomposition (SVD) to find the canonical correlations \citepsm{healy1957rotation, ewerbring1990canonical}. In the SVD method, we firstly find the singular value decompositions of $X$ and $Y$ as
\begin{align*}
    X &= U_X^{} D_X^{} V_X^T\\
    Y &= U_Y^{} D_Y^{} V_Y^T
\end{align*}
where the columns of $U_X^{} (U_Y^{})$ and the columns of $V_X^{} (V_Y^{})$ are called the left and right singular vectors of $X (Y)$, respectively. Then, we can find the canonical correlations by finding the singular values of $U_X^T U_Y^{}$. Overall, this method requires three singular value decompositions. 

Another way to compute canonical correlations is to find the QR decomposition of $X$  and $Y$ and then apply SVD \citepsm{golub1969matrix, bjorck1973numerical}. The QR decomposition of $X$ and $Y$ are
\begin{align*}
    X &= Q_X R_X\\
    Y &= Q_Y R_Y
\end{align*}
where $Q_X^{T} Q_X^{} = I_p$, $Q_Y^{T} Q_Y^{} = I_q$, and $R_X$ and $R_Y$ are upper triangular matrices. Following this, we can express $\Sigma_X$ and $\Sigma_Y$ as
\begin{align*}
    \Sigma_X &= X^T X^{} \\
    &= R_X^T Q_X^T Q_X^{} R_X^{} \\
    &= R_X^T R_X^{} \\
    \Sigma_Y &= Y^T Y^{} \\
    &= R_Y^T Q_Y^T Q_Y^{} R_Y^{} \\
    &= R_Y^T R_Y^{}
\end{align*}
where $\Sigma_X = R_X^T R_X^{}$ and $\Sigma_Y = R_Y^T R_Y^{}$ are the Cholesky decompositions. The singular values of $Q_X^T Q_Y^{}$, which are derived from
\begin{align*}
    (R_X^T)^{-1} \Sigma_{XY}^{} R_Y^{-1} &= (R_X^T)^{-1} R_X^T Q_X^T Q_Y^{} R_Y^{} R_Y^{-1}\\
    &= Q_X^T Q_Y^{},
\end{align*}
are the canonical correlations. This method requires two QR decompositions and one singular value decomposition. Since QR decomposition for a matrix requires less computational work than computing SVD \citepsm{do2012numerically}, we utilise QR decomposition to compute canonical correlations
in the implementation of our methods. 

\section{Variable importance} 

In this section, we explain computation of the variable importance for covariates in random forest framework and present the simulation results for the estimated variable importance. 

\subsection{Variable importance computation} \label{supp:vimpexplain}

The quantification of variable importance (VIMP) is important to assess relative importance of covariates. In the random forest framework, the mostly used VIMP idea proposed by \citesm{breiman2001random} is based on the increase in the prediction error when the link between the covariates and the response is broken by permuting out-of-bag (OOB) observations. In a regression setting with a set of covariates $X$ and for a continuous response variable $Y$, VIMP for $X_j$ can be computed as
\begin{equation*}
    VIMP(X_j) = \frac{1}{B} \sum_{b=1}^{B} \big(MSE(OOB_j^b) - MSE(OOB^b)\big)
\end{equation*}
where $OOB^b$ is the OOB sample of the $b^{th}$ tree of the forest, $OOB_j^b$ is the OOB sample of the $b^{th}$ tree where the $j^{th}$ covariate is randomly permuted and $MSE$ stands for the mean squared error. The average over $B$ trees gives the variable importance measure for $X_j$. Larger VIMP shows greater importance. 

\texttt{randomForestSRC} uses the same idea of breaking the link between covariates and the response but with another way of permuting $X_j$ as proposed in \citesm{ishwaran2007variable}. Instead of permuting OOB samples at each tree, during the tree growing process, observations in the parent node are assigned to child nodes at random or consistently to the other child node when the split variable is $X_j$. VIMP for $X_j$ can be computed as
\begin{equation} \label{eq:vimprfsrc}
    VIMP(X_j) = \frac{1}{B} \sum_{b=1}^{B} \big(MSE(\tilde{t}^b, OOB^b) - MSE(t^b, OOB^b)\big)
\end{equation}
where $t^b$ is the original $b^{th}$ tree of the forest, $\tilde{t}^b$ is the permuted $b^{th}$ tree, and $OOB^b$ is the OOB sample of the $b^{th}$ tree. 

Computing VIMP measure \eqref{eq:vimprfsrc} requires the true response values for training observations. However, in this paper the problem is unsupervised by nature. Therefore, we propose a two-step process for VIMP computation. Firstly, we build a random forest with the proposed splitting criterion for $X$, $Y$, $Z$ and compute the estimated canonical correlations, $\hat \rho(z_i)$, as described in Algorithm 1 in the paper. Then, we use the $\hat \rho(z_i)$ estimations as a continuous response variable for the original covariates ($Z$) and we train a regression random forest. Finally, we use the VIMP measures from this random forest. 

\subsection{Simulation results for variable importance} \label{supp:vimpresult}
\begin{figure}[H]
    \centerline{\includegraphics[width=\textwidth]{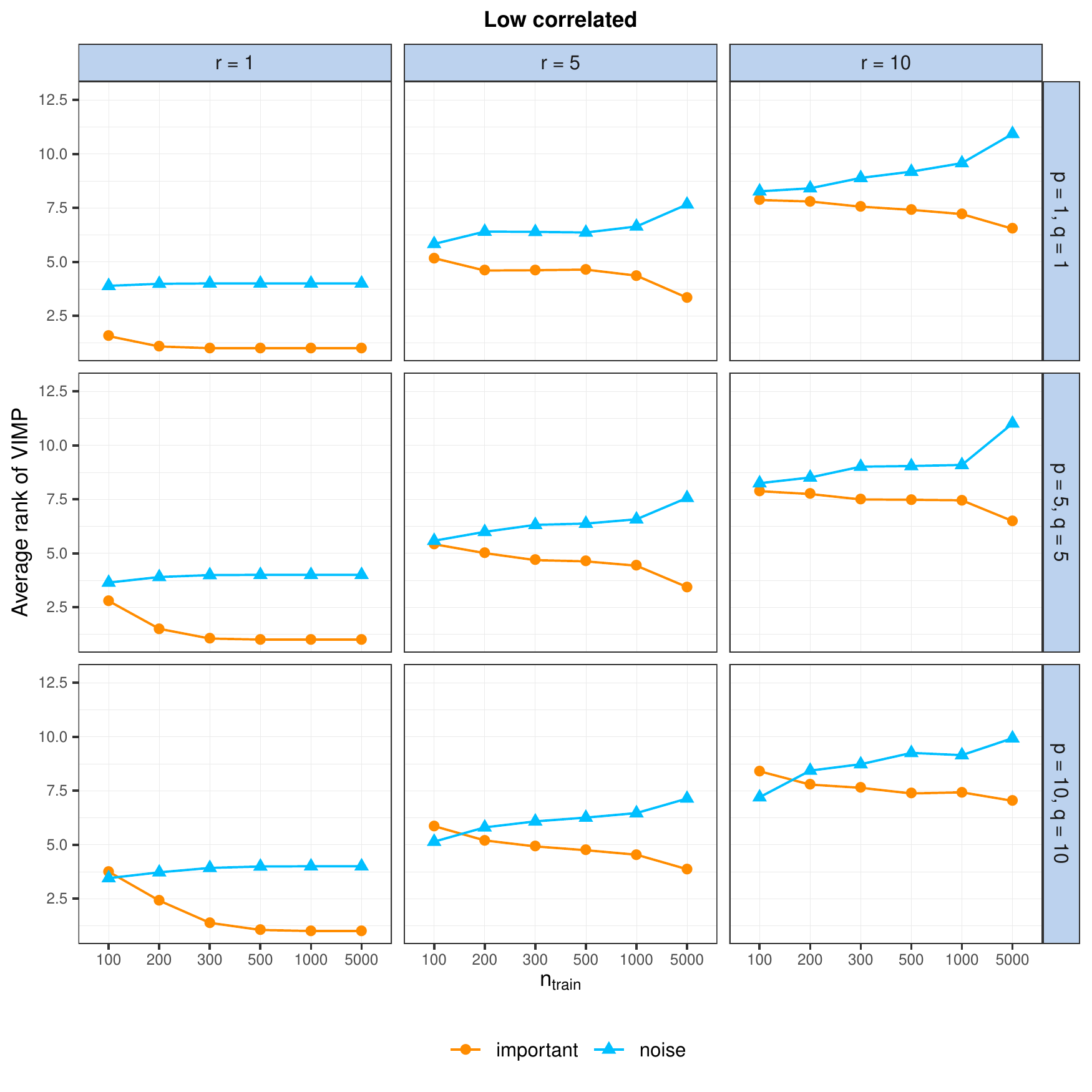}}
    \caption{Average ranks from estimated VIMP measures for low correlated data sets when \texttt{nodesize} = $3\times (p+q)$. Smaller values of rank indicate a more important variable (the most important variable has rank 1).}
    \label{fig:vilowcor3}
\end{figure}

\begin{figure}[H]
    \centerline{\includegraphics[width=\textwidth]{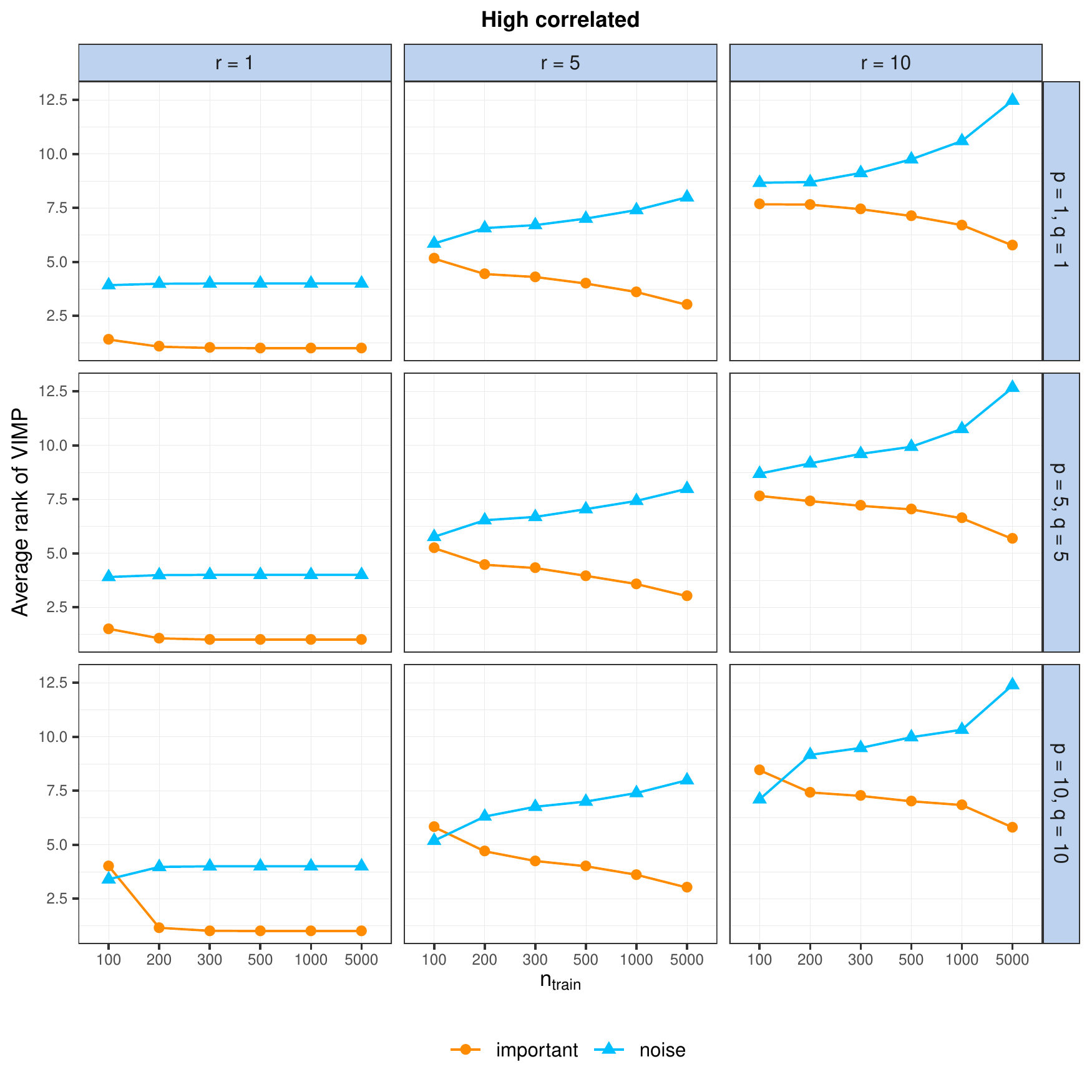}}
    \caption{Average ranks from estimated VIMP measures for high correlated data sets when \texttt{nodesize} = $3\times (p+q)$. Smaller values of rank indicate a more important variable (the most important variable has rank 1).}
    \label{fig:vihighcor3}
\end{figure}

As stated in Section \ref{subsubsec:vimp} of the paper, Figures \ref{fig:vilowcor3} and \ref{fig:vihighcor3} present the average rank, from the estimated VIMP measures, for the important variables group and noise variables group, for low and high correlated data sets, respectively. The most important variable (the one with the highest VIMP measure) has rank 1. As ranks increases, variable importance decreases. We see that in almost all settings, the important variables have
smaller average ranks than noise variables.

\section{Examples of sample distributions with DGP} \label{supp:dgp}

\begin{figure}[H]
    \centerline{\includegraphics[width=\textwidth]{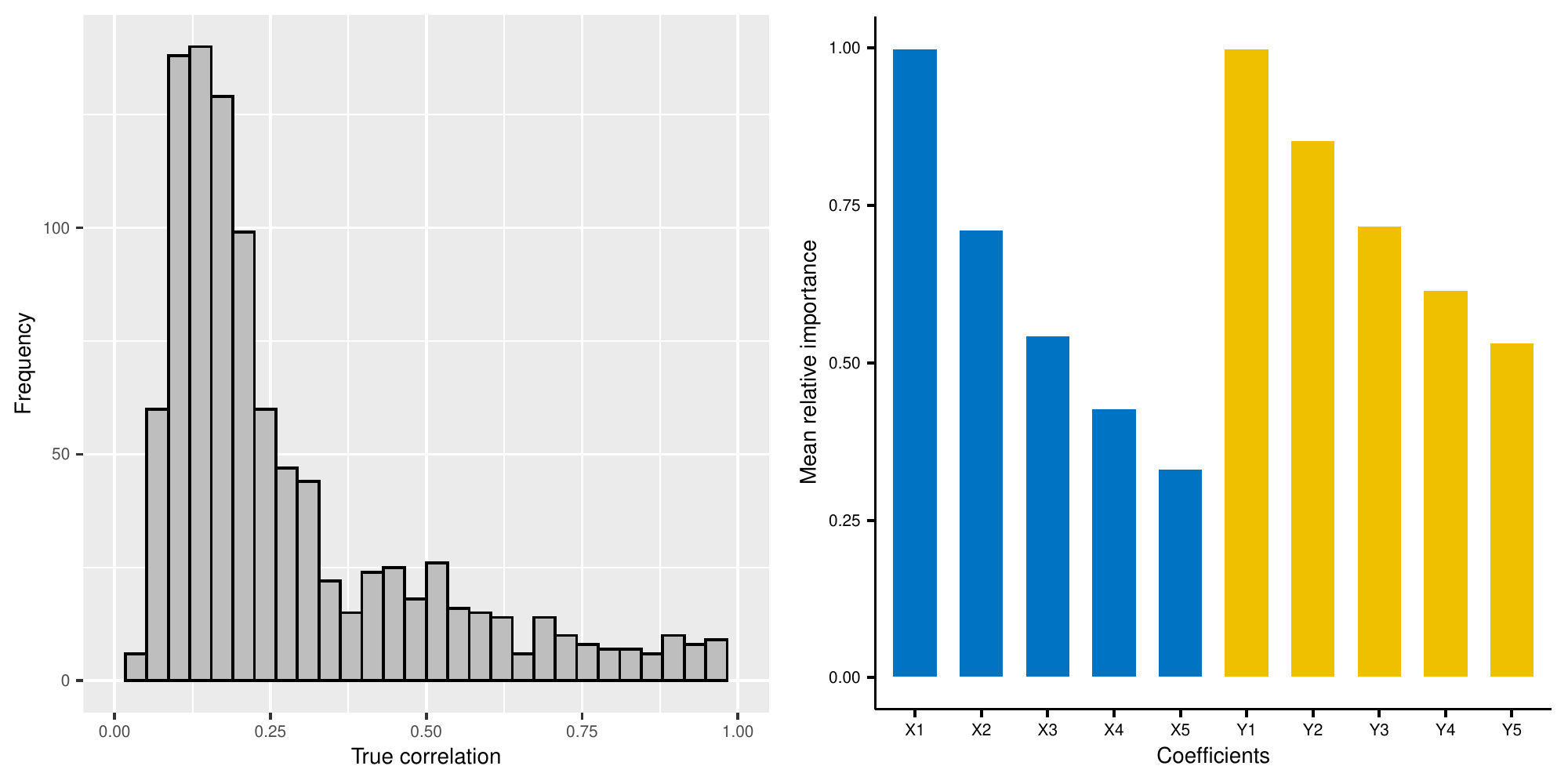}}
    \caption{In this example, we have $5X, 5Y, 5Z$ variables and the sample size is 1000. The DGP parameters are $\beta_0 = -2$, $\rho_x = \rho_y = 0.3$, $\rho_z = 0.1$, $s_x = 0.7$, $s_y = 0.4$. The mean and median correlations are 0.29 and 0.20, respectively. This setting corresponds to the low correlated data set with $n_{train} = 1000$. \textit{(left)} The sample distribution of correlations. \textit{(right)} The ordered bar chart for the average of the coefficients within the $X$ and $Y$ sets over the sample.}
    \label{fig:example1}
\end{figure}

\begin{figure}[H]
    \centerline{\includegraphics[width=\textwidth]{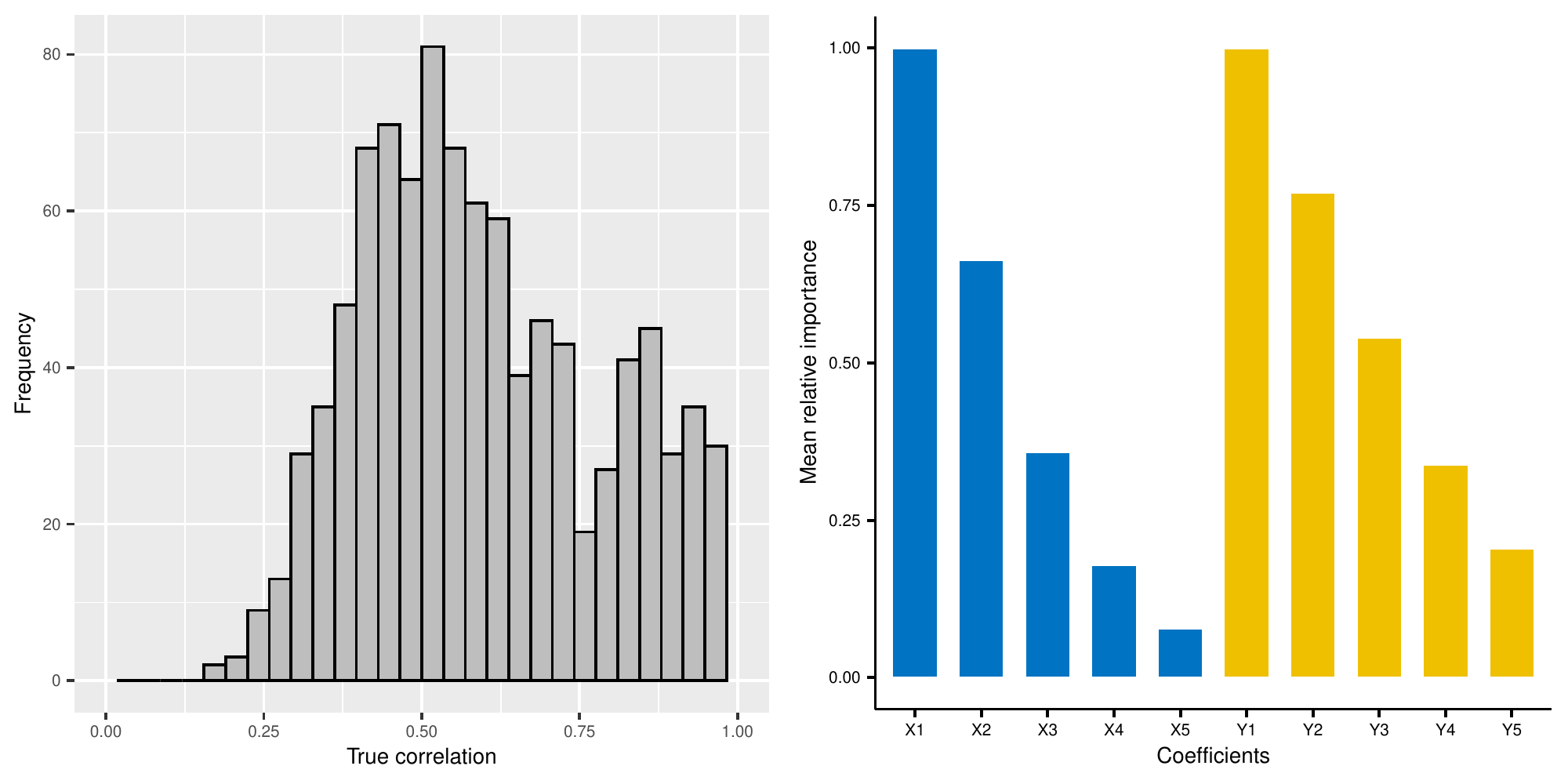}}
    \caption{In this example, we have $5X, 5Y, 5Z$ variables and the sample size is 1000. The DGP parameters are $\beta_0 = -0.3$, $\rho_x = \rho_y = 0.3$, $\rho_z = 0.1$, $s_x = 0.4$, $s_y = 0.3$. The mean and median correlations are 0.61 and 0.57, respectively. This setting corresponds to the high correlated data set with $n_{train} = 1000$. \textit{(left)} The sample distribution of correlations. \textit{(right)} The ordered bar chart for the average of the coefficients within the $X$ and $Y$ sets over the sample.}
    \label{fig:example2}
\end{figure}

Figures \ref{fig:example1} and \ref{fig:example2} show some examples for sample distributions with different parameter settings. See Section \ref{subsec:dgp} of the paper for the explanation of DGP. Figure \ref{fig:example1} and \ref{fig:example2} correspond to the low and high correlated data settings with $n_{train} = 1000$ in the simulations for accuracy evaluation (Section \ref{subsubsec:accuracyevaldesign} of the paper), respectively. The left plot in the figures is the histogram of the generated sample. In the low correlated data setting (Figure \ref{fig:example1}), the mean and median correlations of the sample are 0.29 and 0.20, respectively. In the high correlated setting (Figure \ref{fig:example2}), the mean and median correlations of the sample are 0.61 and 0.57, respectively. The right plot in the figures is the ordered bar chart for the average of the coefficients within the $X$ and $Y$ sets over the sample. The selection of parameters $s_x$ and $s_y$ affects the coefficients of variables. In both low and high correlated settings, $s_x > s_y$ which result in faster decrease in generated $X$ coefficients ($a$) compared to $Y$ coefficients ($b$). 

\section{Simulations results for accuracy evaluation} \label{supp:nodesizecomparison}

In this section, we present the detailed simulations results for accuracy evaluation. See Section \ref{subsubsec:accuracyevaldesign} of the paper for the explanation of performance criterion.

\subsection{Results for \texttt{nodesize} selection}

\begin{figure}[H]
    \centerline{\includegraphics[width=0.8\textwidth]{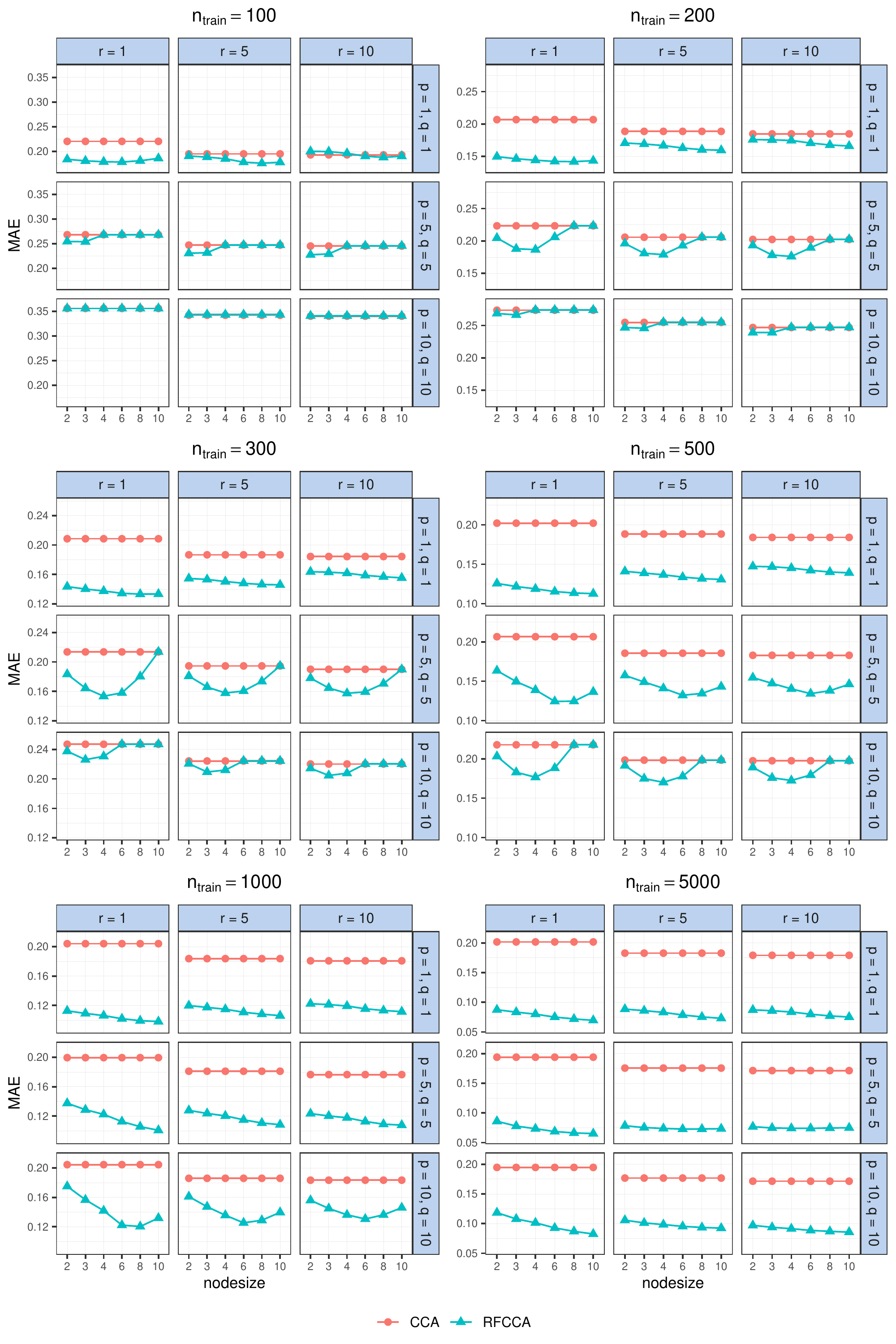}}
    \caption{Accuracy evaluation results for low correlated data sets. $r^{noise} = 5$ in all settings. The values in the \textit{x}-axis correspond to the levels of \texttt{nodesize} parameter. CCA is the benchmark method. Smaller values of $MAE$ are better.}
    \label{fig:lowcornodesize}
\end{figure}

\begin{figure}[H]
    \centerline{\includegraphics[width=0.8\textwidth]{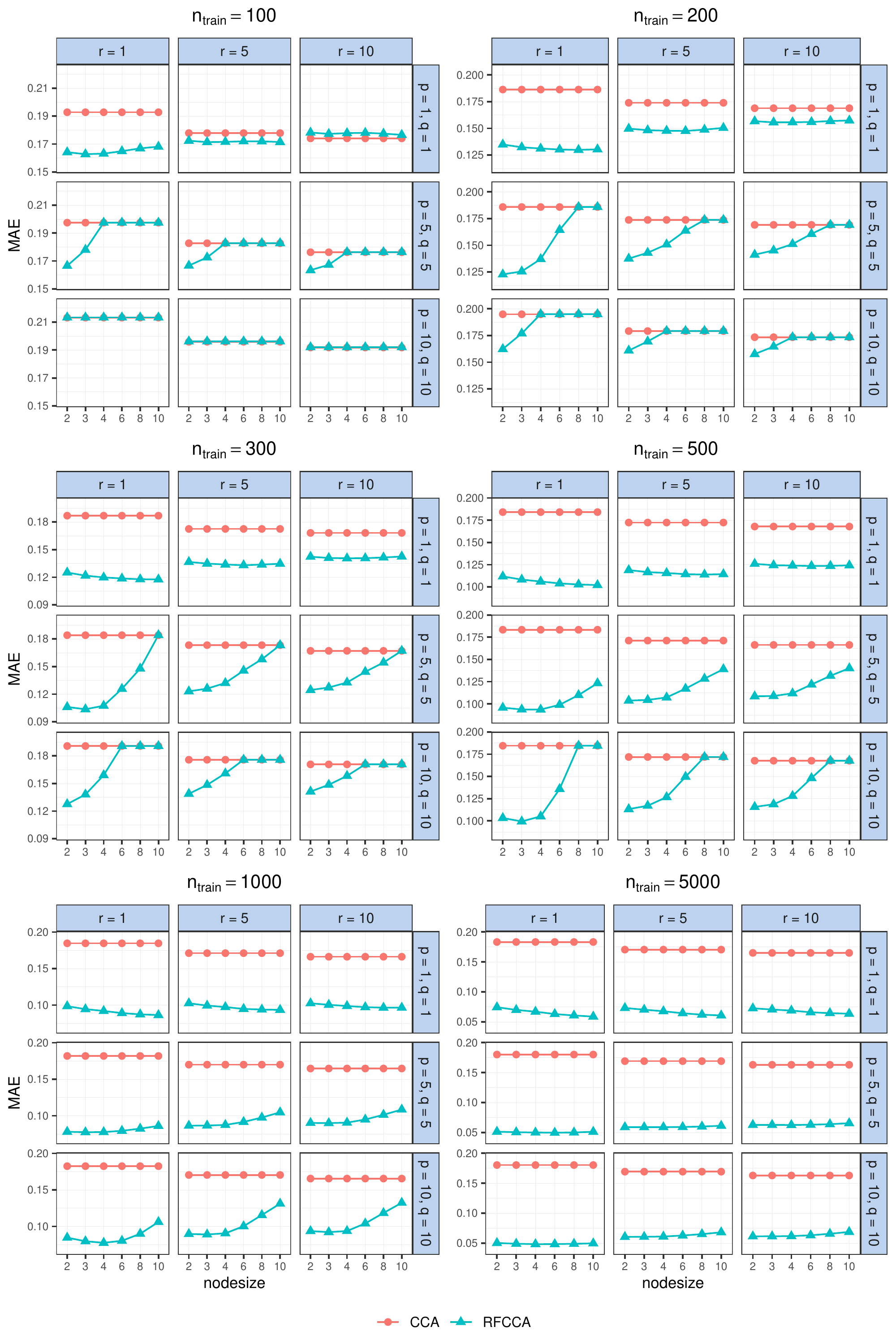}}
    \caption{Accuracy evaluation results for high correlated data sets. $r^{noise} = 5$ in all settings. The values in the \textit{x}-axis correspond to the levels of \texttt{nodesize} parameter. CCA is the benchmark method. Smaller values of $MAE$ are better.}
    \label{fig:highcornodesize}
\end{figure}

Here we compare the performance of the proposed method with six levels of \texttt{nodesize}, $\{2\times (p+q), 3\times (p+q), 4\times (p+q), 6\times (p+q), 8\times (p+q), 10\times (p+q)\}$ , for low and high correlated data settings and $n_{train}=\{100,200,300,500,1000,5000\}$. Figures \ref{fig:lowcornodesize} and \ref{fig:highcornodesize} present the average $MAE$ over the 100 repetitions with each $n_{train}$ for the low and high correlation settings, respectively. As can be seen from the results, in some scenarios, $MAE$ increases as \texttt{nodesize} increases (\textit{e.g.}, high correlation setting with $p=10,q=10,r=5$ and $n_{train}=1000$) whereas in some scenarios $MAE$ decreases as \texttt{nodesize} increases (\textit{e.g.}, low correlation setting with $p=5,q=5,r=5$ and $n_{train}=1000$). Moreover, there are some cases where $MAE$ decreases first and then increases as \texttt{nodesize} increases (\textit{e.g.}, low correlation setting with $p=5,q=5,r=5$ and $n_{train}=300$). In such situations, normally we do a hyperparameter tuning with cross-validation or by dividing the data set into train and validation sets. However, in our case, we cannot tune the \texttt{nodesize} parameter because we do not have a target. As can be seen from the \texttt{nodesize} comparison results, although it is not optimal, the proposed method is almost always better than the benchmark method. 

\begin{figure}[H]
    \centerline{\includegraphics[width=\textwidth]{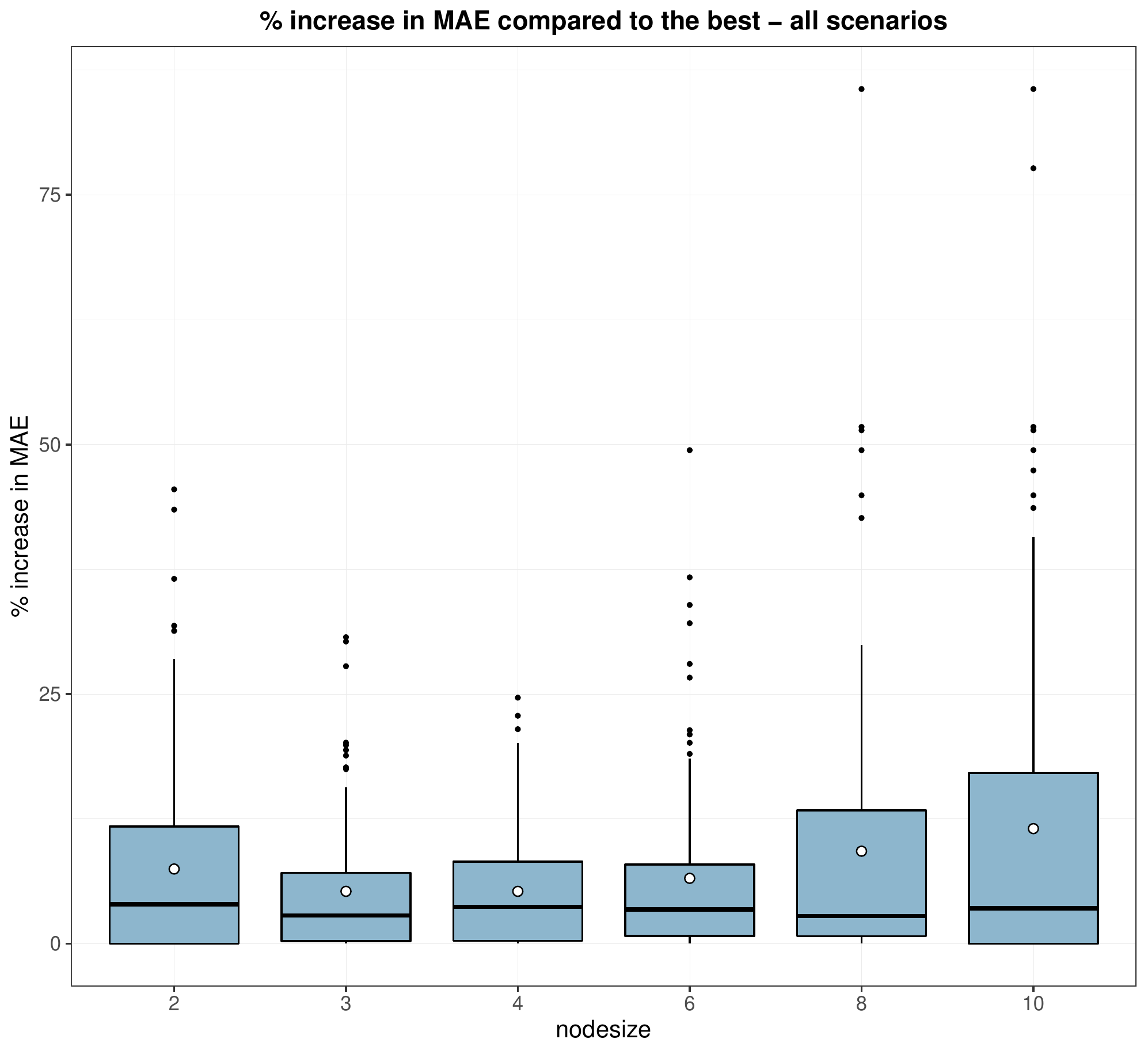}}
    \caption{Distributions of the percentage increase in MAE of each level of \texttt{nodesize} parameter with respect to best performing \texttt{nodesize} value for a given scenario for all 108 scenarios in the simulation study.}
    \label{fig:nodesizeallscenarios}
\end{figure}

The best accuracy is achieved with different levels of \texttt{nodesize} parameter for each scenario and it is hard to select the global best level of \texttt{nodesize} for all scenarios from those figures. Hence, we provide a global view of the performance for \texttt{nodesize} parameter in Figure \ref{fig:nodesizeallscenarios}. To be able to compare the accuracy of the proposed method with different levels of \texttt{nodesize} parameter across different scenarios, we use the percentage increase in $MAE$ with respect to the best performer \texttt{nodesize} value for a given scenario. This way, we can aggregate the results across all scenarios. There are 108 scenarios for each level of \texttt{nodesize} parameter in this simulation study (2 mean CCA correlation levels $\times$ 3 $Z$ dimensionality $\times$ 3 $X$ and $Y$ dimensionality $\times$ 6 training sample sizes). For a given scenario, we have the $MAE$ on the test set for each \texttt{nodesize} level. Let $MAE_{n}$ be the $MAE$ of \texttt{nodesize} $n$, where $n=\{2\times (p+q), 3\times (p+q), 4\times (p+q), 6\times (p+q), 8\times (p+q), 10\times (p+q)\}$, for this scenario. The percentage increase in $MAE$ of \texttt{nodesize} $n$ with respect to the best performer for this scenario is computed as
$$100 \times \frac{MAE_n - min_n\{MAE_n\}}{min_n\{MAE_n\}}$$
where $min_n\{MAE_n\}$ is the smallest $MAE$ for this scenario. Hence, the smaller values indicate better accuracy. Figure \ref{fig:nodesizeallscenarios} shows the distributions of this measure for the \texttt{nodesize} parameter. The results show that, the mean and median of $3 \times (p+q)$, $4 \times (p+q)$ and $6 \times (p+q)$ are very similar and among them $3 \times (p+q)$ has a smaller median and interquartile range. Hence, in the paper, we evaluate the accuracy of the proposed method by setting \texttt{nodesize} $=3 \times (p+q)$ for all scenarios.

\newpage
\subsection{Performance results for different values of parameter \texttt{nodesize}}

\begin{figure}[H]
    \centerline{\includegraphics[width=0.8\textwidth]{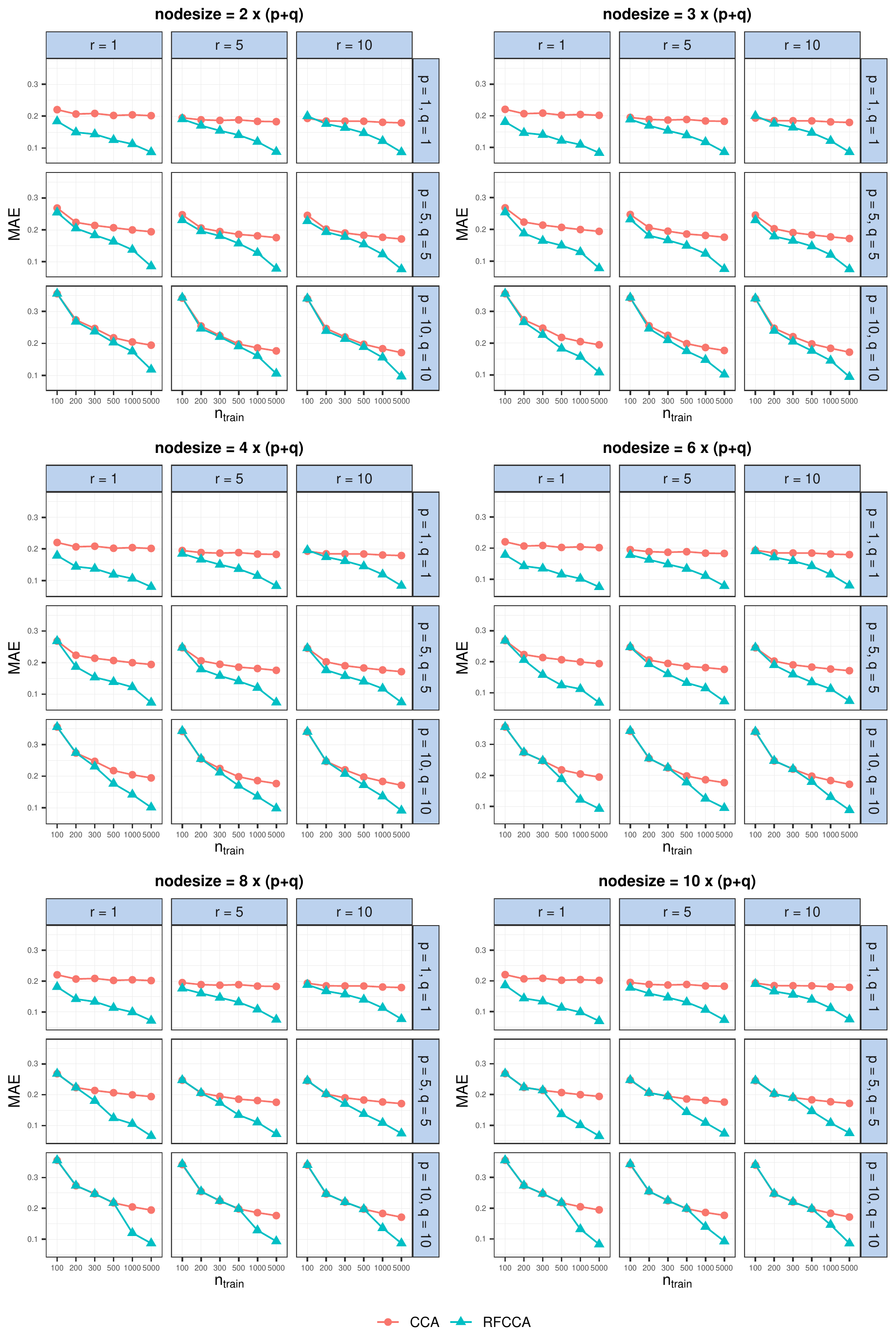}}
    \caption{Accuracy evaluation results for low correlated data sets for six values of parameter \texttt{nodesize} = $\{2\times (p+q), 3\times (p+q), 4\times (p+q), 6\times (p+q), 8\times (p+q), 10\times (p+q)\}$. $r^{noise} = 5$ in all settings. CCA is the benchmark method. Smaller values of MAE are better.}
    \label{fig:lowcorntrain}
\end{figure}

\begin{figure}[H]
    \centerline{\includegraphics[width=0.8\textwidth]{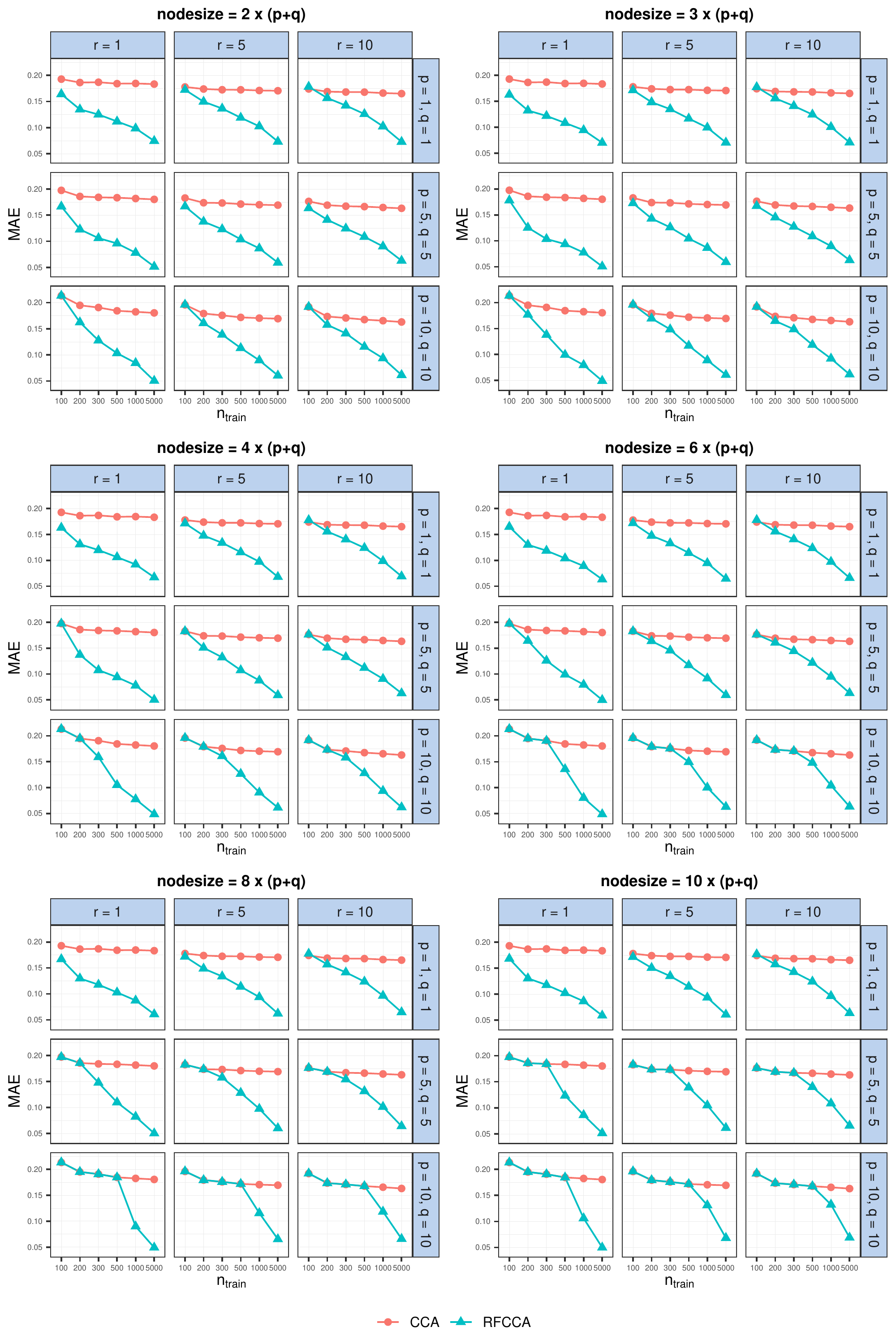}}
    \caption{Accuracy evaluation results for high correlated data sets for six values of parameter \texttt{nodesize} = $\{2\times (p+q), 3\times (p+q), 4\times (p+q), 6\times (p+q), 8\times (p+q), 10\times (p+q)\}$. $r^{noise} = 5$ in all settings. CCA is the benchmark method. Smaller values of MAE are better.}
    \label{fig:highcorntrain}
\end{figure}

Figures \ref{fig:lowcorntrain} and \ref{fig:highcorntrain} present the average $MAE$ over the 100 repetitions when \texttt{nodesize} = $\{2\times (p+q), 3\times (p+q), 4\times (p+q), 6\times (p+q), 8\times (p+q), 10\times (p+q)\}$ for the low and high correlation settings, respectively. In fact, top right plot in Figure \ref{fig:lowcorntrain}, which shows the results of \texttt{nodesize} = $3\times (p+q)$, is the Figure \ref{fig:lowcor3} in the paper and is represented here to be able to compare with the results of other \texttt{nodesize} values. Similarly, top right plot in Figure \ref{fig:highcorntrain} is the Figure \ref{fig:highcor3} in the paper. As can be seen in Figures \ref{fig:lowcorntrain} and \ref{fig:highcorntrain}, for the larger \texttt{nodesize} values $MAE$ of the proposed method and the benchmark method are the same for scenarios with smaller $n_{train}$. For example, when \texttt{nodesize} = $6\times (p+q)$, $MAE$ of the proposed method and the benchmark method are the same for $p=10, q=10$ when $n_{train} = \{100,200,300\}$. For this setting, \texttt{nodesize} $= 6\times(10+10)=120$ which results in no splits for $n_{train} = \{100,200\}$. For $n_{train} = 300$, a single split can occur in trees with this \texttt{nodesize}. However, the $BOP$ of a new observation may include all of the training observations due to randomness. Each tree of the forest is a stump and there is a high chance that the union of the training observations that are in the same terminal nodes as the new observation is equal to the set of training observations. Estimating canonical correlation with this $BOP$ is the same as computing CCA for all $X$ and $Y$. When the sample size is small, increasing the \texttt{nodesize} may cause underfitting. 

\subsection{Performance results for bootstrapping and sub-sampling} \label{supp:samplingcomparison}

\begin{figure}[H]
    \centerline{\includegraphics[width=\textwidth]{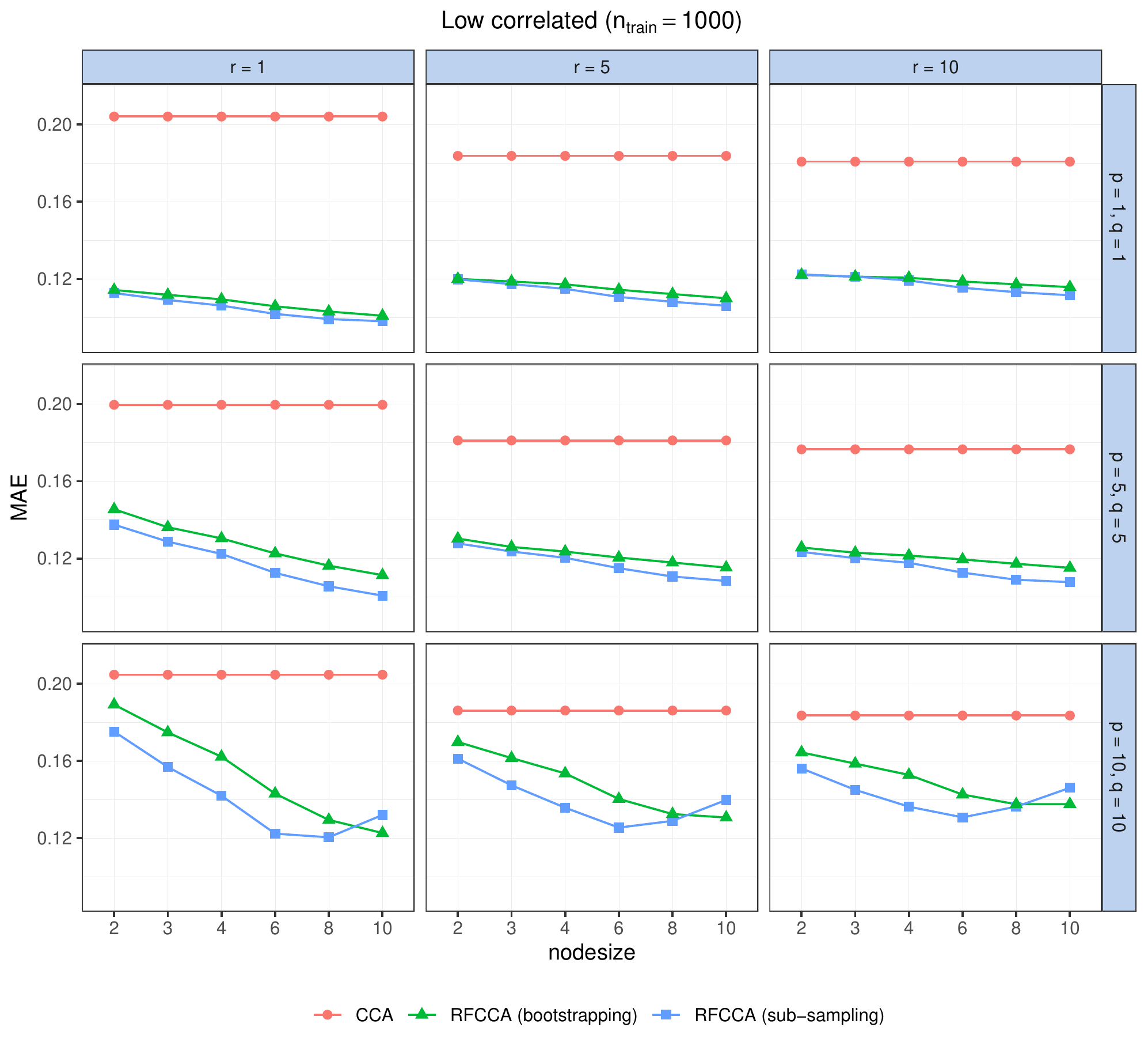}}
    \caption{Accuracy evaluation results for low correlated data sets when $n_{train}=1000$. $r^{noise} = 5$ in all settings. CCA is the benchmark method. Smaller values of MAE are better.}
    \label{fig:lowcorsampling}
\end{figure}

\begin{figure}[H]
    \centerline{\includegraphics[width=\textwidth]{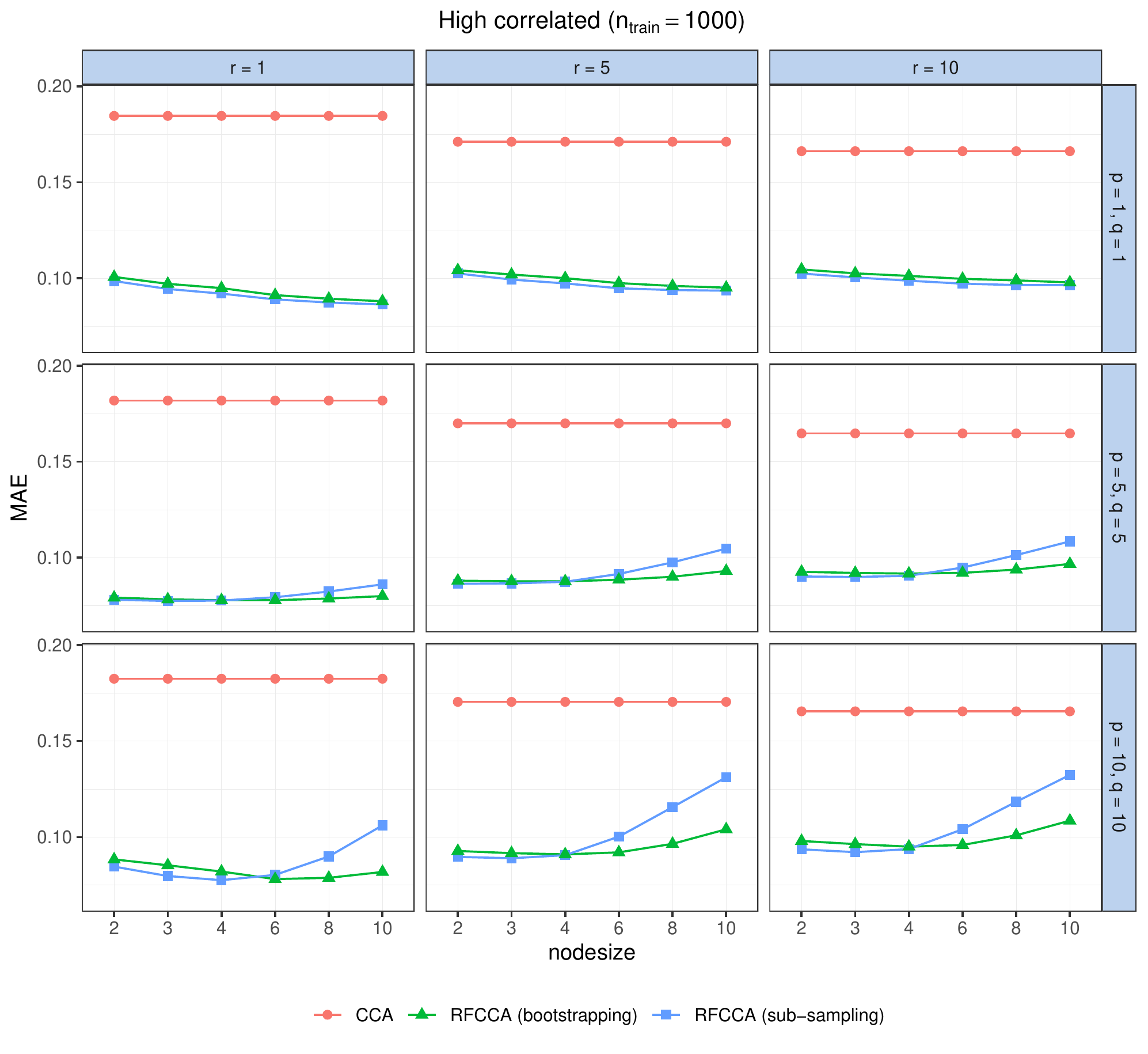}}
    \caption{Accuracy evaluation results for high correlated data sets when $n_{train}=1000$. $r^{noise} = 5$ in all settings. CCA is the benchmark method. Smaller values of MAE are better.}
    \label{fig:highcorsampling}
\end{figure}

Here we investigate the effect of bootstrapping and sub-sampling on the performance of the proposed method. Figures \ref{fig:lowcorsampling} and \ref{fig:highcorsampling} present the average $MAE$ over the 100 repetitions when $n_{train}=1000$ for the low and high correlation settings, respectively. The values in the \textit{x}-axis correspond the values of the \texttt{nodesize} parameter which are $\{2\times (p+q), 3\times (p+q), 4\times (p+q), 6\times (p+q), 8\times (p+q), 10\times (p+q)\}$. We can both compare the effect of sampling method and \texttt{nodesize} parameter on the accuracy with those plots. CCA is used as the benchmark method. In most of the settings, there is no significant difference in performance between sub-sampling and bootstrapping. However, in some cases (\textit{e.g.} low correlated data sets with $p=10,q=10$), sub-sampling has slightly better accuracy than bootstrapping. There are also some cases, for instance in high correlated data sets with $p=10,q=10$, sub-sampling shows better performance for the smaller \texttt{nodesize} values whereas bootstrapping has smaller $MAE$ for the larger \texttt{nodesize} values. However, in those cases the best accuracy is still obtained with sub-sampling and smaller \texttt{nodesize} values. Hence, sub-sampling is recommended. Overall, although the optimal value for the \texttt{nodesize} parameter may change with the selected sampling method, the accuracy of the proposed method with both sampling methods have a very similar pattern for different levels of \texttt{nodesize} parameter.

\newpage

\section{EEG data} \label{supp:eeg}
In this section, the data collection process and preprocessing steps are explained.

\subsection{Neuropsychological scale}
To evaluate the intellectual disabilities, performance IQ (pIQ) and verbal IQ (vIQ), which are the scores for total IQ, were obtained using the Mullen, the WPPSI-IV, the WISC-V, and the WAIS-IV batteries depending on the age of the participant. The administration was adapted for the clinical participants by starting at the first item rather than the starting point for their chronological age, frequent breaks were proposed, and participants were motivated through many creative incentives if necessary (songs, games, conversations about their interests, etc.).

\subsection{EEG auditory task}
The auditory task was prepared using E-prime 2.0 software (Psychology Software Tools Inc., Pittsburgh, PA, USA) on a screen placed at a viewing distance of 60 cm. Sounds, which are presented binaurally and simultaneously, were delivered through two speakers located laterally at 30 cm from the participants’ ears. The auditory stimulus consisted of 24 dB/octave white noise burst. Each stimulus lasted 50 ms with an inter-stimulus interval varying between 1200 and 1400 ms to avoid a process of habituation. The task was composed of 150 trials. The total task duration was around 4 minutes. To assure maximal collaboration, a movie without sound and subtitles was presented. The participant was told to focus his/her attention on the movie and not to give attention to the auditory stimuli.

\subsection{EEG recordings}
The subject was placed in an electrically shielded room in the Sainte Justine’s Hospital. The continuous EEG was recorded with a high-density EEG system containing 128 electrodes placed according to the extended 10 - 20 system (Electrical Geodesics System Inc., Eugene, OR, USA). Signals were acquired and processed by a G4 Macintosh computer using NetStation EEG Software (Version 2.0). Before recording, impedances were verified and were kept below 40 k$\Omega$ \citepsm{tucker1993spatial}. EEG data were amplified, band-pass-filtered 0.1--4000 Hz, and sampled at 1000 Hz with a vertex reference.

\subsection{Preprocessing for the analysis}
Off-line signal processing and ERP analyses were performed using the EEGLAB toolbox via custom Matlab scripts \citepsm{delorme2004eeglab}. EEG acquired data was subjected to the following pre-processing steps. First, EEG signals were digitally filtered with a high-band pass filter (0.5 Hz) and a 60 Hz notch filter. Twenty-eight electrodes placed on the neck and the face and containing muscular artifacts have been removed to avoid contamination of average reference. Moreover, a voltage threshold method (2--200 $\mu$V) was applied to exclude channels with artifacts. Data were off-line re-referenced to the mean of the EEG selected electrodes (100 channels). Independent component analysis (ICA) as implemented in the EEGLAB toolbox (with default parameters) was used to remove ocular artifacts. By removing or minimizing the effects of overlapping components, ICA enables a detailed examination of the separate dynamics of electrical brain activity as well as artifacts to remove them \citepsm{delorme2007enhanced}. Ocular and cardiac ICA components (range across subjects: 1--3 components) were identified by visual inspection and deleted from the global signal. Continuous EEG was segmented into epochs covering a time window from -200 ms to 800 ms relative to the onset of the tone. Also, as the artifact ICA components could be deemed unsatisfactory, the segmented EEG recordings were visually inspected by a well-trained experimenter, and trials presenting with residual artifacts were rejected. 

For time-frequency (TF) and inter-trial coherence (ITC) analyses, segments were exported to MATLAB (version R20174b) (The MathWorks Inc., Natik, MA, USA) after artifact rejection. TF and ITC analyses were processed using the EEGLAB toolbox (v.13.6.5b) (La Jolla, CA, USA). TF analysis allows us to explore different frequency bands in terms of their power and temporal distributions \citepsm{herrmann200511}. We used complex Morlet’s wavelet transformation \citepsm{tallon1999oscillatory} to provide power maps in time and frequency domains. The simplified mathematical expression for applying this specific wavelet convolution on our EEG signal is as follows:
\begin{equation*}
    M(t,f) = \int_{t} W\Big(\frac{t-a}{b}, f\Big) S(t) dt
\end{equation*}
where $M(t,f)$ is a matrix of complex values (vectors) for a given time ($t$) and frequency ($f$), $S$ is the signal as a function of time ($t$) and $W$ corresponds to Morlet’s wavelet which is a complex exponential (Fourier) with a Gaussian envelope that undergoes a series of translations ($a$) and dilations ($b$) dependently of the frequency ($f$). The event-related spectral perturbation (ERSP) computation uses the complex values (amplitude and phase) given by Morlet’s wavelet transform as shown in the following formula calculating the power spectrum for each time and frequency point: 
\begin{equation*}
   P(t,f) = 10 \log_{10} |M(t,f)|^2
\end{equation*}
where $P(t,f)$ denotes TF power in terms of decibels (dB). Final TF maps were computed as follows:
\begin{equation*}
   TF = \frac{1}{N} \sum_{n=1}^{N} P(t,f)
\end{equation*}
where $N$ is the total number of trials. The range of frequency investigated was from 3 to 100 Hz. 
ITC, analogous to phase-locking value (PLV), allows the assessment of the strength of phase coherence across trials in temporal and spectral domains \citepsm{makeig2004mining}. The ITC computation uses only the phase of the complex values given by Morlet’s wavelet transform. ITC was computed as in \citesm{lachaux1999measuring} to extract PLV. ITC measures phase coupling across trials at all latencies and frequencies and is defined by: 
\begin{equation*}
   ITC = \frac{1}{N} \Big|\sum_{n=1}^{N} exp(j\theta(f,t,n))\Big|
\end{equation*}
where $\theta$ represents the phase for a given frequency ($f$), time point ($t$), and trial ($n$). The obtained values are always defined between 0 and 1. Phase-locking values close to 1 indicate strong inter-trial phase-locking, thus representing evoked activity while scores closer to 0 indicate a high inter-trial phase variability, thus representing induced activity \citepsm{lachaux1999measuring}.

\subsection{SHAP interaction effect between age and vIQ}
\begin{figure}[H]
    \centerline{\includegraphics[width=0.5\textwidth]{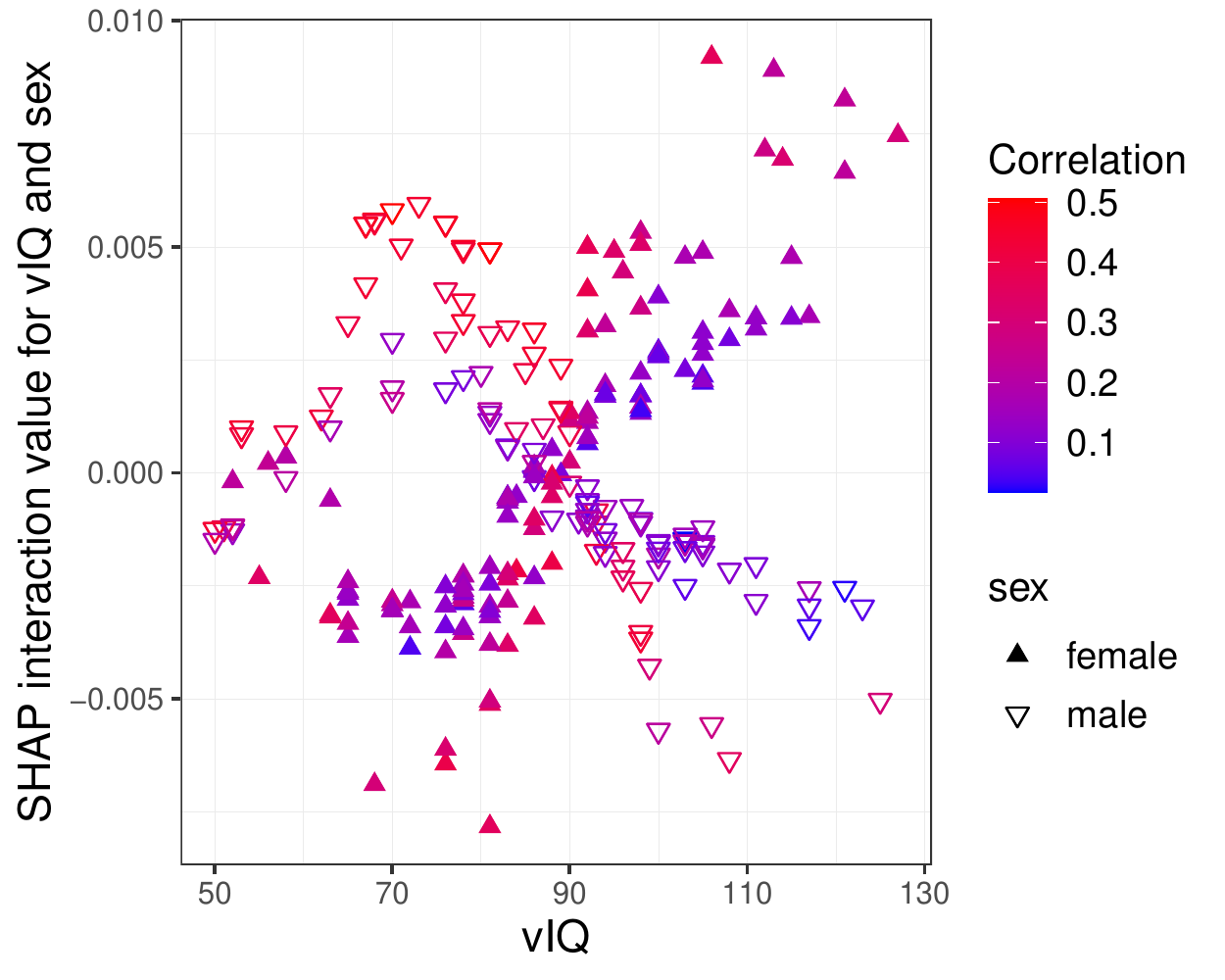}}
    \caption{SHAP interaction values for sex and vIQ variable.}
    \label{suppfig:interviq}
\end{figure}

As mentioned in Section \ref{sec:realdata} of the paper, Figure \ref{suppfig:interviq} presents the interaction effect between age and vIQ. Similar to the interaction effect between sex and pIQ (right plot in Figure \ref{fig:shapeffect} of the paper), we see that the impact increases as we move away from the average vIQ. Again, the impact of the interaction on the theta-gamma co-variation is positive for high IQ females and negative for low IQ females whereas the opposite is observed in males.

\section{Conditional inference framework within our context} \label{supp:condinf}

In principle, the proposed permutation test for covariates’ effects could be used to select the split variable at a node,  analogous to the conditional inference framework \citepsm{hothorn2006unbiased}. However, the computational cost would likely be high. More precisely, assume we are at a node and want to decide whether to split it or not and with which covariate. Using only the observations in the node; 
\begin{enumerate}
    \item For one covariate at a time, apply the permutation test described in Section \ref{subsec:globalsignificance} of the main paper. However, permute only the rows of the given covariate instead of permuting rows of covariate set ($Z$). Obtain a \textit{p}-value for that covariate. 
    \item Repeat Step 1 for all covariates to obtain one \textit{p}-value per covariate. 
    \item If none of the covariates are significant (after applying a multiple testing correction if deemed appropriate), then do not split the node.
    \item Otherwise, select the covariate with the smallest \textit{p}-value as the split variable. Find the best split using the proposed split criterion. 
\end{enumerate}

\bibliographystylesm{natbib}
\bibliographysm{refs}

\end{document}